\documentclass[prb,aps,twocolumn,superscriptaddress,floatfix,citeautoscript,longbibliography]{revtex4-2}
\usepackage{graphicx,rotating,subfigure,amsmath,amsfonts,amssymb,delarray,color,hyperref}
\usepackage{dsfont}
\usepackage[T1]{fontenc}
\usepackage{multirow}
\usepackage{comment}

\newcommand{\tr}{\text{tr}}

\def\12{\frac{1}{2}}

\begin{document}
% % \bibliographystyle{natphys2}
%\bibliographystyle{apsrev}
%\bibliographystyle{abbrv}

\title{Entanglement dynamics in the three-dimensional Anderson model}

\author{Yang Zhao}
\email{zhaoyang2017@nwpu.edu.cn}
\author{Dingyi Feng}
\email{fengdingyi@nwpu.edu.cn}
\author{Yongbo Hu}
\author{Shutong Guo}
\affiliation{Shanxi Key Laboratory of Condensed Matter Structures and Properties, School of Physical Science and Technology, Northwestern Polytechnical University, Xi'an 710072, China}
\author{Jesko Sirker}
\email{sirker@physics.umanitoba.ca}
\affiliation{Department of Physics and Astronomy, University of Manitoba, Winnipeg R3T 2N2, Canada}
\affiliation{Manitoba Quantum Institute, University of Manitoba, Winnipeg R3T 2N2, Canada}

\date{\today}

\begin{abstract}
We numerically study the entanglement dynamics of free fermions on a
cubic lattice with potential disorder following a quantum quench. We
focus, in particular, on the metal-insulator transition at a critical
disorder strength and compare the results to the putative many-body
localization (MBL) transition in interacting one-dimensional
systems. We find that at the transition point the entanglement entropy
grows logarithmically with time $t$ while the number entropy grows
$\sim\ln\ln t$. This is exactly the same scaling recently found in the
MBL phase of the Heisenberg chain with random magnetic fields
suggesting that the MBL phase might be more akin to an extended
critical regime with both localized and delocalized states rather than
a fully localized phase. We also show that the experimentally easily
accessible number entropy can be used to bound the full entanglement
entropy of the Anderson model and that the critical properties at the
metal-insulator transition obtained from entanglement measures are
consistent with those obtained by other probes.
\end{abstract}

\maketitle
\section{Introduction}
Entanglement measures have been shown to be a useful tool to
investigate the non-equilibrium dynamics of quantum systems. Because
entangling two previously unentangled regions generally requires the
exchange of some entangled entities, they provide direct insights into the
quasiparticle dynamics of a many-body system
\cite{CalabreseCardyQuench}. For a typical clean quantum lattice
system with short-range interactions, excitations will spread in a
lightcone-like fashion with a finite Lieb-Robinson velocity
\cite{LiebRobinson,BravyiHastings}. This leads to a linear increase 
of the von-Neumann entanglement entropy $S$ following a quench from an
unentangled initial state and ultimately to a saturation at long times
to a value proportional to the volume of the considered region. This
behavior is expected to be modified in systems with disorder. The
spreading of quasiparticles is then no longer ballistic and the
saturation values are reduced as compared to the clean case.

The exact behavior depends on the type and strength of the disorder,
the dimensionality of the system, whether or not the system is
interacting, as well as other microscopic details such as the range of
interactions and the properties of the initial state
\cite{ZnidaricProsen,BardarsonPollmann,Prelovsek2016,AndraschkoEnssSirker,EnssAndraschkoSirker,ZhaoAndraschkoSirker,ZhaoAhmedSirker,ZhaoSirker2019,LukinRispoli}. 
For one-dimensional non-interacting quantum lattice models with
short-range hoppings, any amount of potential disorder is known to lead
to the Anderson localization of all eigenstates
\cite{Anderson58,AbrahamsAnderson,EdwardsThouless,AndersonLocalization}. 
The saturation values of the entanglement entropy after a quench are
then proportional to the localization length. In the two-dimensional
case, potential disorder also always leads to the localization of all
eigenstates. The entanglement dynamics following a quantum quench is,
however, much more interesting than in the one-dimensional
case. Whereas in one dimension an initial sub-ballistic growth of the
entanglement entropy is immediately followed by saturation, an
extended intermediate time regime does exist in the two-dimensional
case for weak disorder.  This is due to the fact that in this case the
localization length can be much larger than the mean free path. In
this regime of weak localization, the entanglement entropy after a
quench grows logarithmically in time \cite{ZhaoSirker2019}. Another
interesting case which has been studied previously are one-dimensional
non-interacting lattice models with off-diagonal (bond) disorder
\cite{IgloiSzatmari,ZhaoAndraschkoSirker}. 
These systems are driven towards an infinite randomness fixed point
characterized by a localization-delocalization transition in the
eigenstate energy
\cite{EggarterRiedinger,BalentsFisher97,Fisher_random,Fisher_random_Ising,Fisher_random_Ising2}. As
a consequence, the entanglement entropy after a quench never saturates
but grows in the thermodynamic limit without bounds as $S\sim\ln\ln
t$ \cite{ZhaoAndraschkoSirker}.

To gain even more insights into quench dynamics based on entanglement
probes, one can make use of the fact that for a system with particle
number conservation the reduced density matrix $\rho$---obtained after
tracing out a part of the system---will have a block structure. The
von-Neumann entanglement entropy can then be split into two
parts \cite{KlichLevitov,WisemanVaccaro,DowlingDohertyWiseman,SchuchVerstraeteCirac,SchuchVerstraeteCirac2,SongFlindt,SongRachel,KieferUnanyan1,KieferUnanyan2,KieferUnanyan3,Bonsignori2019,MurcianodiGiulio,MurcianodiGiulio2,LukinRispoli}
\begin{eqnarray}
\label{Sent}
S &=& -\tr\{ \rho\ln\rho\} \\
&=& -\sum_n p(n)\ln p(n) -\sum_n p(n) \tr\{\rho(n)\ln\rho(n)\}\, .\nonumber 
\end{eqnarray}
The first part is the number entropy $S_N$ and the second part the
configurational entropy $S_{\textrm{conf}}$. The probability to find
$n$ particles in the considered subsystem is denoted by $p(n)$, and
$\rho(n)$ is the block of the reduced density matrix with $n$
particles. Note that Eq.~\eqref{Sent} has a beautiful structure: $S_N$
only depends on the probability distribution of particles while
$S_{\textrm{conf}}$ is the von-Neumann entropy of the block $\rho(n)$
of the reduced density matrix weighted by the probability of finding
$n$ particles in the subsystem. $S_N$ does measure particle
fluctuations while $S_{\textrm{conf}}$ measures the degree of
superposition between different configurations with the same particle
number. Apart from being of theoretical interest, this separation of
the entanglement entropy into two parts is also useful for
experimental studies
\cite{LukinRispoli}. Since $S_N$ is much easier to measure than the total entanglement 
entropy---which typically requires to resort to a full quantum
tomography--- it is intriguing to try to characterize $S$ by $S_N$
alone. For Gaussian systems---whether they are clean or
disordered---it has been recently shown that this is indeed possible
and that the relation $S\sim\exp(S_N)$ holds \cite{KieferUnanyan1}.

Within the last decade, the question whether or not localization is
also possible in interacting systems has received renewed attention
\cite{BaskoAleiner,OganesyanHuse,PalHuse,AndraschkoEnssSirker,EnssAndraschkoSirker,AbaninRev2019,NandkishoreHuse,AltmanVoskReview}. 
Numerical studies of the spin-1/2 Heisenberg chain with local magnetic
fields drawn from a box distribution have been interpreted in terms of
an ergodic to many-body localized (MBL) phase transition at some
finite disorder strength \cite{PalHuse,Luitz1}. One of the hallmarks
of the putative MBL phase is the logarithmic growth of $S(t)$
\cite{ZnidaricProsen,BardarsonPollmann} which is believed to be
entirely driven by the configurational entropy while the number
entropy saturates; there is no transport. Such behavior can be
described by effective models of the MBL phase which consist of
exponentially many conserved charges which are coupled by coupling
constants which decay exponentially with distance
\cite{SerbynPapic,HuseNandkishore}. It is also possible to build phenomenological 
renormalization group flows for the ergodic-MBL transition but those
are based on the assumption that isolating clusters do exist and that
there are no resonances \cite{VoskHusePRX,PotterVasseurPRX,Imbrie2016}. Lately,
however, the stability of the MBL phase in the thermodynamic limit has
been challenged in a number of publications
\cite{SuntajsBonca,SuntajsBonca2,KieferUnanyan2,KieferUnanyan3,SelsPolkovnikov}. 
In particular, it has been found that for the system sizes and times
accessible in exact diagonalizations, the scaling relation
$S\sim\exp(S_N)$ continues to hold even for the strongly disordered
spin-1/2 Heisenberg chain. This suggests that particle transport
continues, albeit at a very slow rate, and that the system might
ultimately always thermalize.

Because of the exponential growth of the Hilbert space with the number
of particles in interacting systems, the system sizes which can be
studied by exact diagonalizations are quite small and definitive
statements about the thermodynamic limit are difficult to come by. In
particular, it has been suggested that the observed growth of the
number entropy in the MBL phase might be transient and related to
being still too close to the transition point \cite{LuitzBarLev}. To
gain further insights into the finite size scaling of $S(t)$ and
$S_N(t)$ near a localization transition is one of our motivations to
investigate the much better understood metal-insulator transition in
the three-dimensional Anderson model from the entanglement angle. In
addition, we believe that the results are interesting in their own
right because the number entropy, in particular, is easily accessible
in cold atomic gas experiments---which have already been used to study
Anderson
localization \cite{OspelkausOspelkaus_Localization,RoatiDerrico,BillyJosse}---and
thus can be a useful probe to test the critical properties at the
transition. This is in contrast to inverse participation ratios of
eigenfunctions which are commonly used in theoretical studies but are
difficult to access experimentally.

The article is organized as follows: the model and methods used to
calculate the various entanglement entropies are introduced in
Sec.~\ref{Model}. Our numerical results for the entanglement growth
after the quench are presented in Sec.~\ref{Results}. In the final
section, we discuss our results in comparison to the ones obtained for
the ergodic-MBL transition in one-dimensional interacting systems and
conclude.

\section{Model and Methods}
\label{Model}
In this paper we consider free spinless fermions on a $L\times L\times L$ square lattice with Hamiltonian 
\begin{equation}\label{Ham}
  H=-t\sum_{\langle i,j\rangle}c^{\dagger}_{i}c_{j}+\sum_{i}D_{i}c^{\dagger}_{i}c_{i}
\end{equation}
at half filling with open boundary conditions. $c_{i}$ and
$c^{\dagger}_{i}$ are fermionic annihilation and creation operators on
site $i$, respectively, and $\langle \cdot ,\cdot\rangle$ denotes
nearest neighbors. The hopping amplitude $t$ is site independent while
the potentials $D_i \in [-D/2,D/2]$ are randomly drawn from a box
distribution. The three-dimensional Anderson model \eqref{Ham} has
been extensively studied by many different
methods \cite{AndersonLocalization,BulkaKramer,SlevinMarkos,VasquezRodriguez1,VasquezRodriguez2}.
Numerical studies have shown, in particular, that the model has a
mobility edge and that all eigenstates become localized for
$D_c\gtrsim 17$. Furthermore, the localization length has been found
to diverge at $D_c$ with a critical exponent $\nu\sim 1.5$, and the
wavefunctions at criticality show multifractal properties.

\begin{figure}
  \centering % Requires \usepackage{graphicx}
  \includegraphics[width=0.8\linewidth,height=0.6\linewidth]{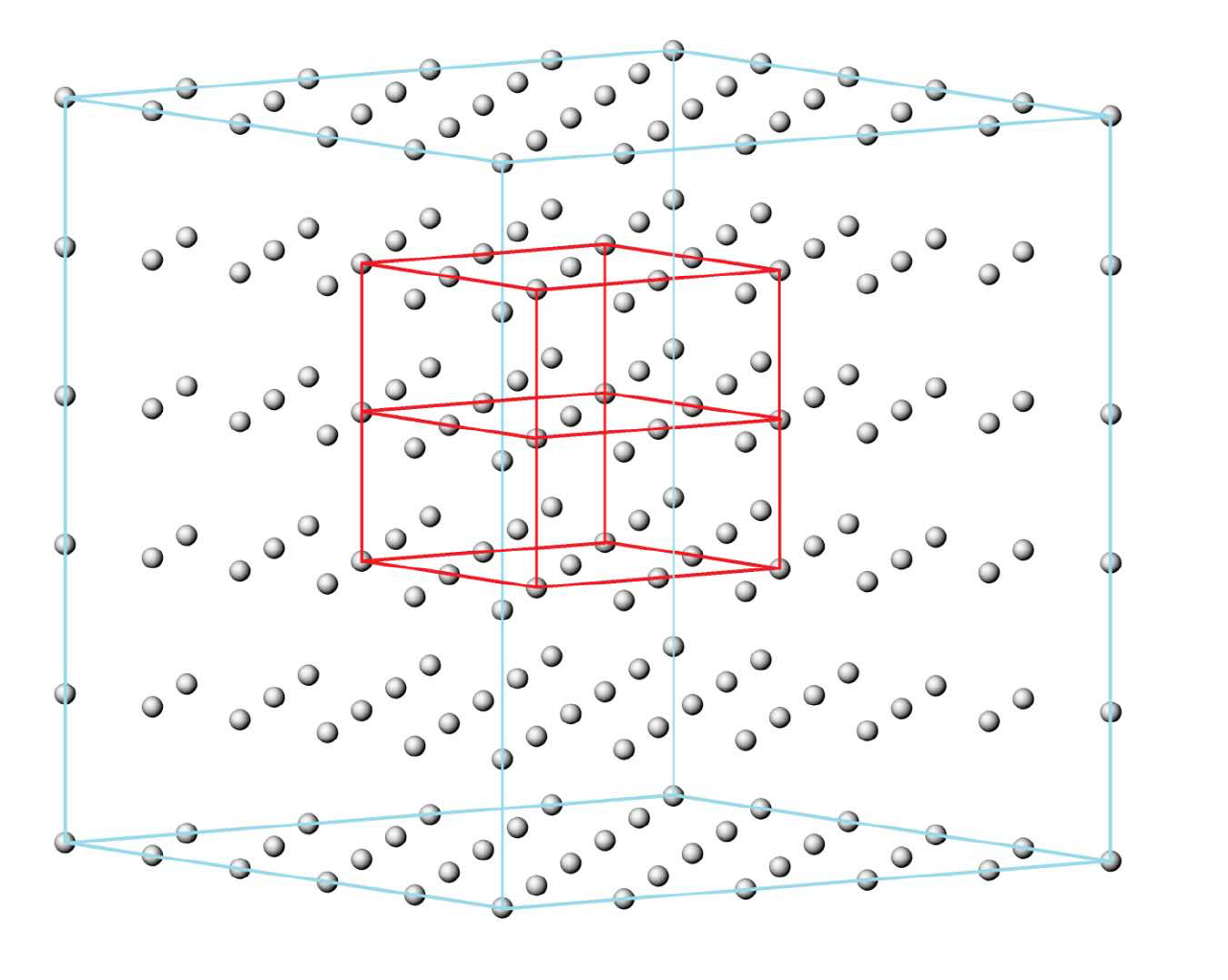}
  \caption{A $6\times 6\times 6$ cubic lattice. The considered
  subsystem is the $3\times 3\times 3$ cube in the middle highlighted
  by the red frame.}
\label{Fig1}
\end{figure}

In the following, we will always prepare the system in the initial
charge density wave (CDW) state
\begin{equation}\label{cdw}
|\Psi(0)\rangle=\prod_{i_1+i_2+i_3\; \textrm{odd}}c_{i_1,i_2,i_3}^{\dagger}|0\rangle,
\end{equation}
where $|0\rangle$ denotes the vacuum state. We have checked explicitly
that none of our conclusions are changed qualitatively if we use
instead random product states as initial states. We then unitarily
evolve the system to obtain the state $|\Psi(t)\rangle
=\exp(-iHt)|\Psi(0)\rangle$ at time $t$. Next, we divide the system
into an inner cube of size
$\frac{L}{2}\times\frac{L}{2}\times\frac{L}{2}$ with the rest of the
cube serving as environment, see Fig.~\ref{Fig1}.
Tracing out the environment we obtain the reduced density matrix,
$\rho(t)=\tr_{\textrm{Env}}|\Psi(t)\rangle\langle\Psi(t)|$, which is
the central object of our study. We are interested in the R\'enyi
entanglement and number entropies of order $\alpha$,
\begin{eqnarray}
\label{Renyi}
S^{(\alpha)} &=& \ln\tr(\rho^\alpha)/(1-\alpha), \\
S_N^{(\alpha)} &=& \ln[\sum_n p^\alpha(n)]/(1-\alpha), \nonumber
\end{eqnarray}
which turn into the von-Neumann entanglement entropy
$S=-\tr\rho\ln\rho$ and the number entropy $S_N = -\sum_n p(n)\ln
p(n)$ in the limit $\alpha\to 1$. As has been shown in
Ref.~\onlinecite{KieferUnanyan1}, the second R\'enyi entropy for
Gaussian systems such as the Anderson model \eqref{Ham} can be bounded by the second
R\'enyi number entropy
\begin{equation}
\label{bound}
\frac{1}{e\pi}\exp\left(2S_N^{(2)}\right)-\frac{1}{6}\leq S^{(2)}\leq \frac{\ln 2}{\pi}\exp\left(2S_N^{(2)}\right) \, .
\end{equation}
This relation will turn out to be useful in our analysis of the
numerical data.

In order to obtain the various entanglement entropies, we use the fact
that for Gaussian systems the eigenvalues $\zeta_i$ of the correlation
matrix $\mathbf{C}$ of the subsystem with matrix elements $C_{nm} =
\langle c^\dagger_n c_m\rangle$ completely determine
its reduced density matrix
\cite{ChungPeschel,Peschel2004,PeschelEisler}. In particular, the
von-Neumann entanglement entropy is given by
\begin{equation}\label{EE}
S=-\sum_i\left(\zeta_i\ln\zeta_i+(1-\zeta_i)\ln(1-\zeta_i)\right),
\end{equation}
and the second R\'enyi entropy can be obtained as
\begin{equation}\label{Renyi2}
S^{(2)}=-\sum_i\ln\left(1-2\zeta_i(1-\zeta_i)\right).
\end{equation}
To calculate the number entropy, we need to find the probability
distribution $p(n)$. This distribution can be obtained by evaluating
the Fourier transformation of the momentum generating function
$\chi(l)$\cite{KlichLevitov,KieferUnanyan1}:
\begin{equation}\label{pn}
p(n)=-\frac{1}{N+1}\sum^N_{l=0}\mathrm{exp}\left(-i\frac{2\pi ln}{N+1}\right)\chi(l),
\end{equation}
with
\begin{equation}\label{pn2}
\chi(l)=-\prod_m\left[1+\left(\mathrm{exp}\left(\frac{i2\pi l}{N+1}\right)-1\right)\zeta_m\right],
\end{equation}
and $N=(L/2)^3$.

Using the formulas above we are able to study the entanglement
dynamics in the three-dimensional Anderson model \eqref{Ham} for
linear system sizes of up to $L=22$ by exact diagonalization. We
average the obtained results for the entropies over at least $200$ and
up to $10,000$ disorder realizations.

\section{Results}
\label{Results}
In order to investigate the effects of the on-site disorder potential
on the entanglement entropy growth following a quantum quench, we
consider disorder potentials from $D=0$ (no disorder) up to $D=200$, thus
moving from the ballistic regime all the way to the strongly localized
regime.

\subsection{Free spinless fermions without disorder}
\begin{figure}
	\centering % Requires \usepackage{graphicx}
	\includegraphics[width=0.9\linewidth]{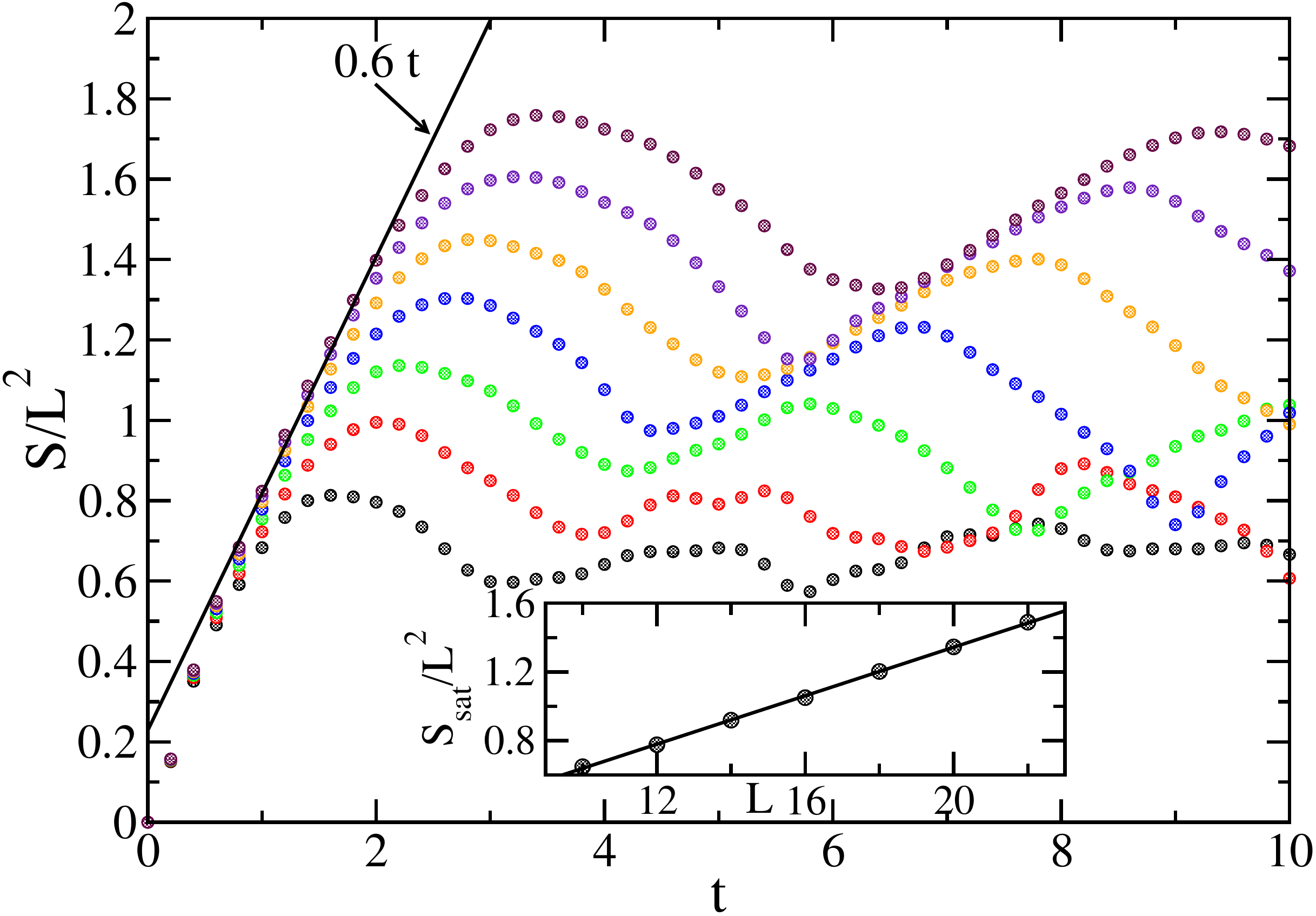}
	\caption{Entanglement entropy $S(t)$ for free fermions without
	disorder on cubic lattices with linear dimensions $L=10$ to
	$L=22$ (bottom to top). Inset: Averaged saturation values as
	function of $L$ and a volume-law fit $S_{sat}\sim 0.07L^3$.}
\label{fg2}
\end{figure}

For fermions with short-range hoppings and no disorder, we expect quite
generally that a quantum quench---starting from a product
state---generates quasi-particle excitations which move ballistically
through the lattice with some maximum velocity $v$ whose scale is set
by the bandwidth\cite{CalabreseCardyQuench}. For the chosen geometry,
this will lead to a linear increase of entanglement up to times
$t\lesssim \ell/4v$ where $\ell = L/2$ is the linear size of the
subsystem. At long times, we expect an average saturation value
following a volume law, $S_{\textrm{sat}}\sim L^3$. Our numerical
simulations confirm this picture, see Fig.~\ref{fg2}.
We do see, in particular, that for small times all system sizes fall
onto a single curve $S(t)/L^2\sim 0.6t$ confirming the linear increase
in time. We note that the effective velocity is smaller than in the
one-dimensional case \cite{ZhaoAndraschkoSirker}, $S(t)\sim 0.88 t$,
but comparable to the one found for the two-dimensional square
lattice \cite{ZhaoSirker2019}, $S(t)/L\sim 0.7t$. The saturation
values, around which $S(t)$ oscillates at long times, are well fitted
by a volume law, see the inset of Fig.~\ref{fg2}.

\begin{figure}
	\centering % Requires \usepackage{graphicx}
	\includegraphics[width=1.0\linewidth]{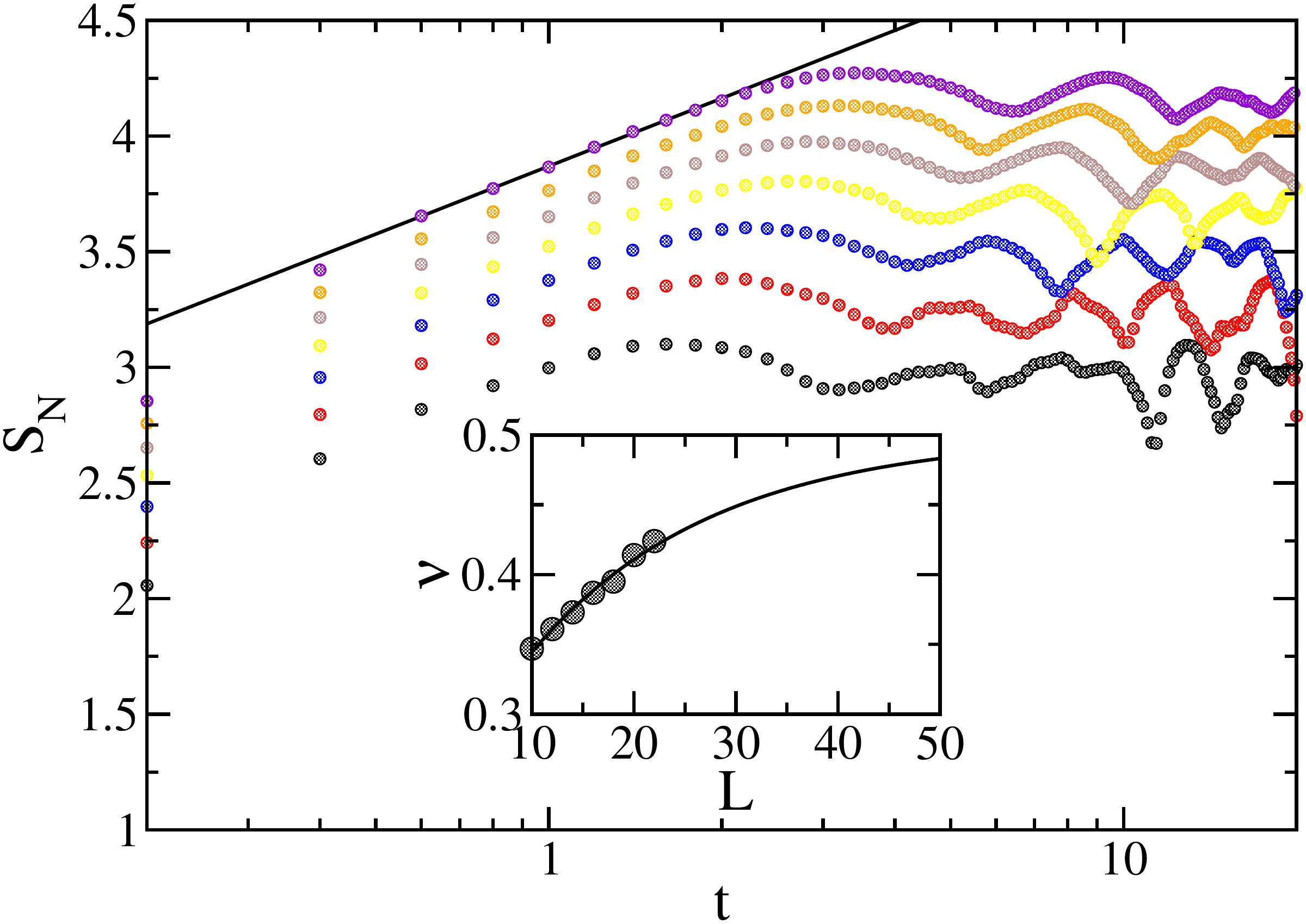} 
\caption{Number entropy for the system without disorder for system sizes $L=10,12,\cdots,22$ (from bottom to top). The line is a fit $S_N=a+\nu\ln t$. Inset: Fit parameters $\nu$ as a function of system size and a fit $\nu=0.5-b\exp(-\gamma L)$.}
\label{fg3}
\end{figure}

Next, we decompose the von-Neumann entropy according to
Eq.~\eqref{Sent} into number and configurational entropy, see
Fig.~\ref{fg3}.
In the thermodynamic limit we expect $S_N\sim \nu\ln t$ with
$\nu=1/2$ \cite{KieferUnanyan1}. The numerical data are consistent with
this expectation, see the inset of Fig.~\ref{fg3}. The finite size
corrections however are still substantial even for $L=22$ and the
convergence to $\nu=1/2$ is slow. We note that the linear growth of
the total entanglement entropy is driven by the configurational
entropy. The number entropy is a much smaller, sub-leading
contribution.

\begin{figure}
	\centering % Requires \usepackage{graphicx}
	\includegraphics[width=1.0\linewidth]{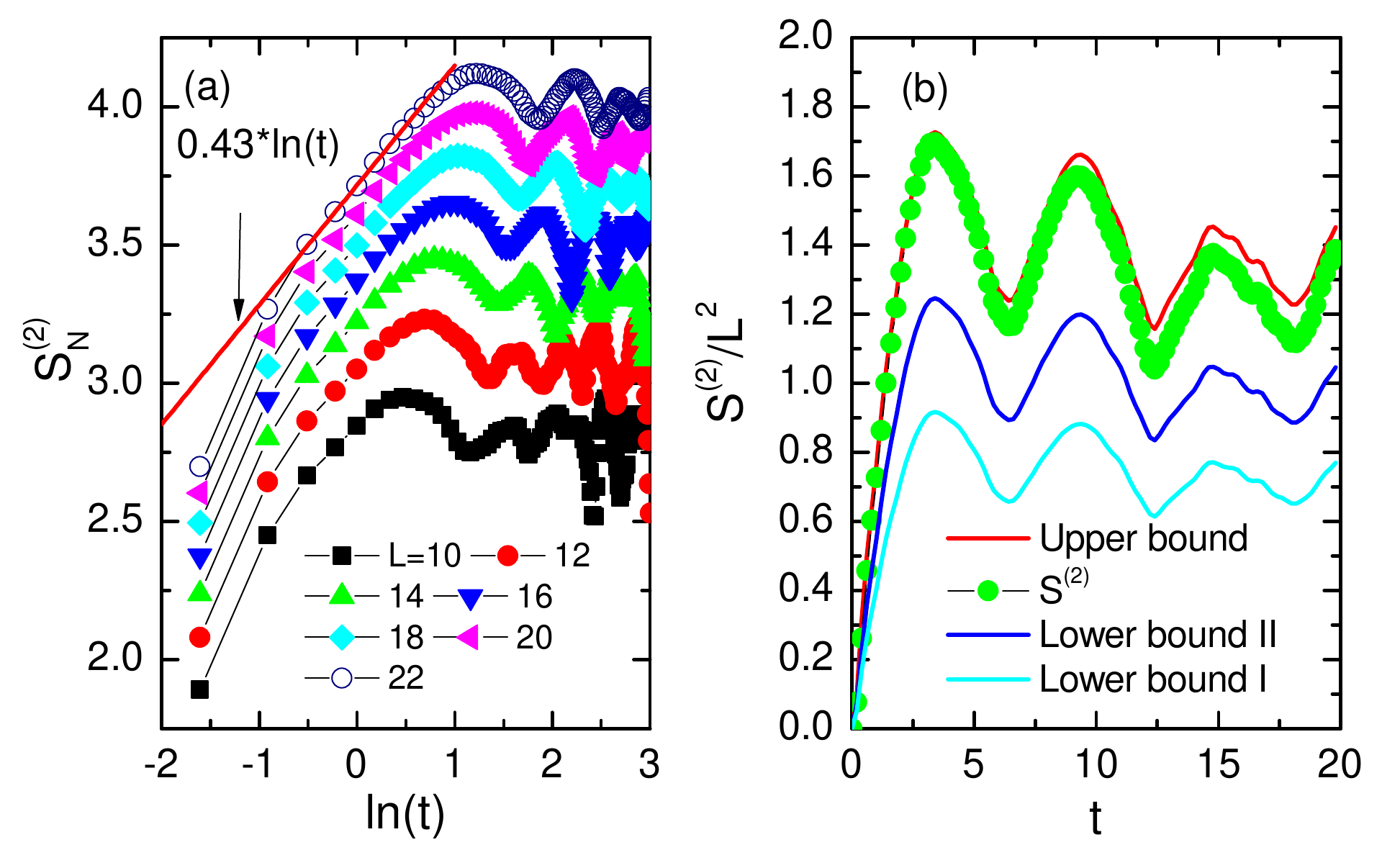} \caption{(a)
	The second R\'enyi number entropy $S^{(2)}_N$ for the system
	without disorder for different $L$. For $L=22$, the scaling
	region is fitted by $S^{(2)}_N\sim 0.43\ln(t)$. (b) The second
	R\'enyi entropy with its upper and two lower bounds,
	$\exp(2S_N^{(2)})/(\text{e}\pi)-1/6$ (Lower bound I) and
	$\exp(2S_N)/(\text{e}\pi)-1/6$ (Lower bound II).} \label{fg4}
\end{figure}

We also consider the decomposition of the second R\'enyi entropy into
number and configurational entropy, see Fig.~\ref{fg4}.
One can show that the scaling $S^{(\alpha)}_N\sim \frac{1}{2}\ln t$
holds for all $\alpha$ in a free fermionic system without disorder
\cite{KieferUnanyan1,KieferUnanyan2}. The numerical data are in good
agreement with this prediction, see Fig.~\ref{fg4}(a), with
finite-size corrections similar to $S_N(t)$ as shown in
Fig.~\ref{fg3}. Finally, the quality of the bounds \eqref{bound} is
checked in Fig.~\ref{fg4}(b). As in the one-dimensional case
considered in Ref.~\onlinecite{KieferUnanyan1}, the upper bound is a
fairly tight bound without disorder. More generally, it is expected
that the upper bound is quite tight if the quasi-particle excitations
spread ballistically while the lower bound is expected to become
better in the strongly localized regime.

\subsection{Weak disorder and criticality}
\begin{figure}
	\centering % Requires \usepackage{graphicx}
	\includegraphics[width=0.9\linewidth]{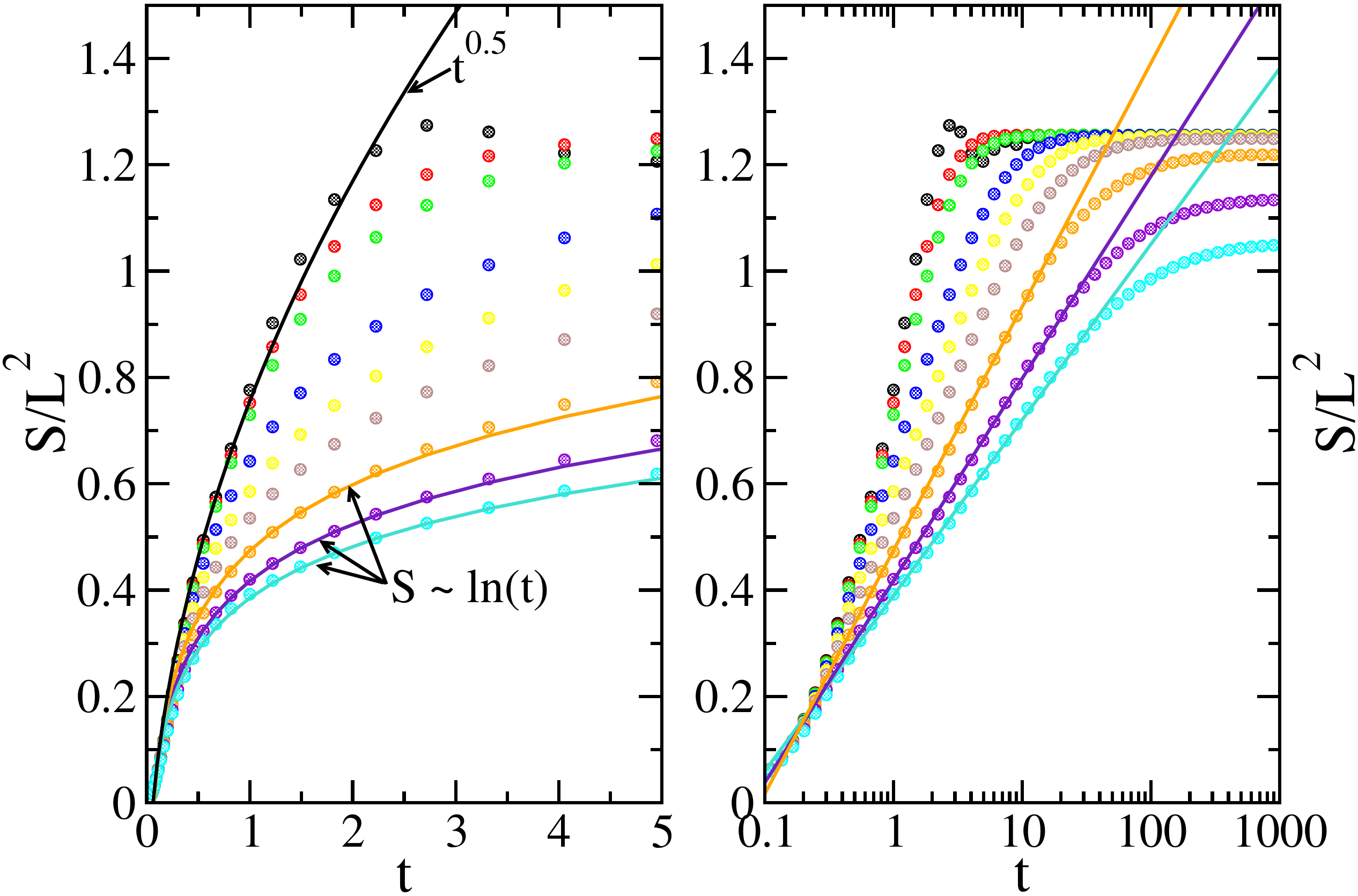} \caption{The
	von-Neumann entropies of lattices with $L=16$ and
	$D=2,4,5,8,10,12,15,18,20$ (top to bottom) on a linear time
	scale (left panel) and a logarithmic time scale (right
	panel). Averages over $1000$ disorder realizations are
	shown.}\label{fg5}
\end{figure}

For the three-dimensional Anderson model it is known that
eigenstates--- depending on their energy---are starting to become
localized at $D_c\approx 16.5$. The critical regime is characterized
by a mobility edge where extended and localized eigenstates coexists
and by a multifractal scaling of the inverse participation ratios of
these states \cite{VasquezRodriguez1,VasquezRodriguez2}. For $D<D_c$ we
are in the metallic phase and expect diffusive behavior. For the
entanglement entropy this means that we expect $S/L^2\sim
\sqrt{t}$. This behavior is consistent with the numerical data
for $L=16$ and $D\lesssim 5$, see Fig.~\ref{fg5}.
For $5\lesssim D\lesssim D_c$ we see that $S(t)$ deviates from a
$\sqrt{t}$ behavior fairly quickly due to finite-size effects. As we
will show exemplarily in Fig.~\ref{fg7} for $D=16$, the data in this
regime can be fitted by an effective power law $S(t)\sim t^\beta$ with
an exponent which increases with system size and is expected to
converge to $\beta=1/2$ in the thermodynamic limit. Physically, this
is consistent with the fact that with increasing disorder fewer and
fewer paths exist which allow for diffusion so that larger and larger
system sizes are needed to observe it over a long time scale. It is
thus clear that one has to be careful in interpreting the finite-size
data: From simply fitting the data for a particular system size one
might come to the conclusion that the power-law exponent in the
entropy growth is decreasing below $1/2$. However, such sub-diffusive
behavior is not expected to be present in this model and the data
indeed show that the effective exponent increases towards $1/2$ with
increasing system size for all $D<D_c$.

For disorder strengths in the critical regime, we find that the data
are well described by $S(t)\sim\ln t$. As shown in the right panel of
Fig.~\ref{fg5}, the data are consistent with a logarithmic growth over
almost three orders of magnitude in time. This behavior can be
understood as follows: For a state at critical energy $E_c$, the
entanglement entropy will scale as $S/L^2\sim \ln
L$ \cite{JiaSubramaniam}. Furthermore, we know that the localization
length is given by $\xi(E)\sim |E-E_c|^{-\nu}$ where $\nu$ is a
critical exponent with $\nu\sim 1.5-1.6$ for the three-dimensional
Anderson model \cite{BulkaKramer,SlevinMarkos}. In the thermodynamic
limit we might replace $\ln L$ by $\ln \xi(E)$, average over energy
and then replace the energy scale (bandwidth) by inverse time
resulting in $S(t)/L^2\sim\ln t$.

\begin{figure}
	\centering % Requires \usepackage{graphicx}
	\includegraphics[width=0.9\linewidth]{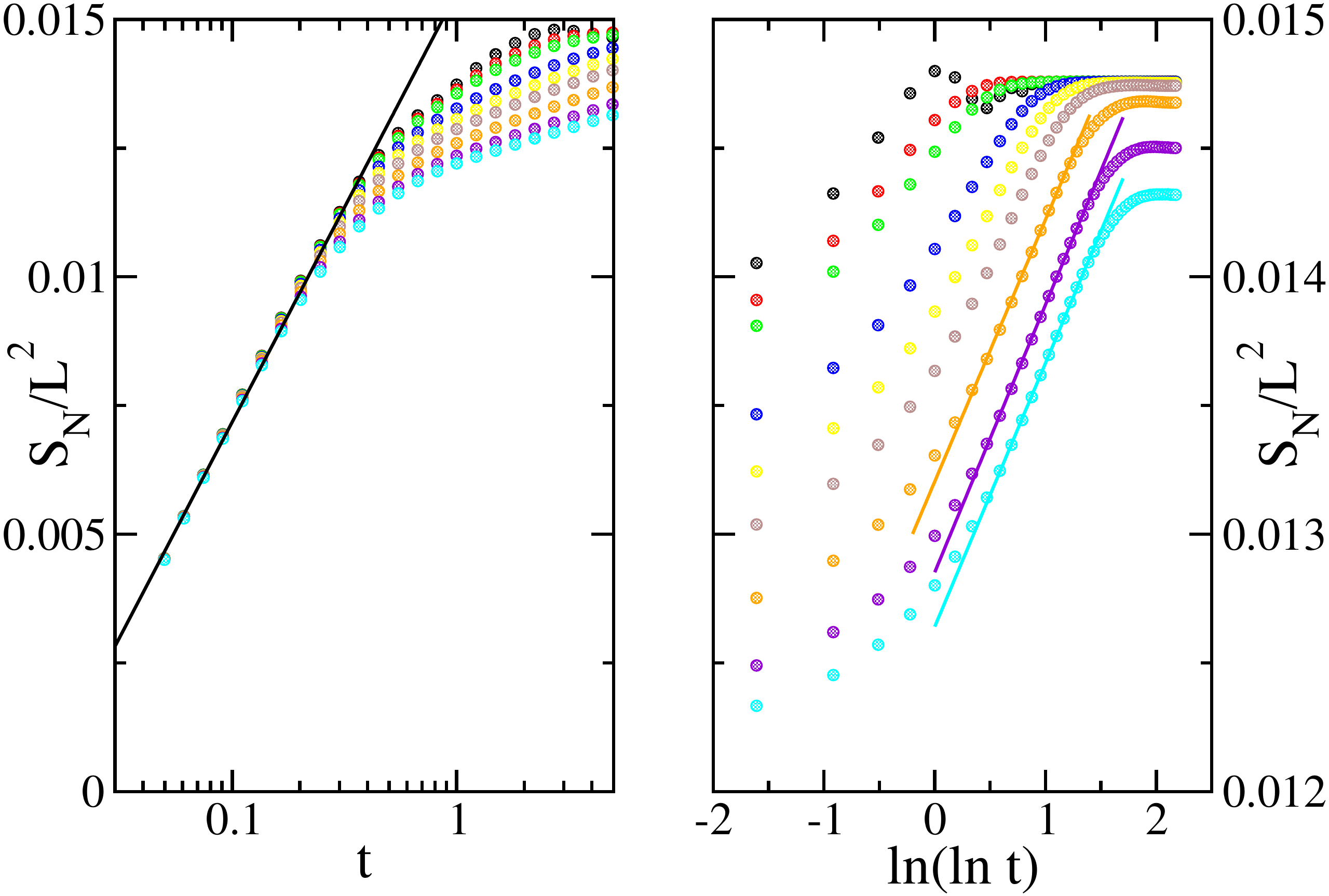} \caption{The
	number entropies corresponding to the entanglement entropies
	shown in Fig.~\ref{fg5} on a logarithmic (left panel) and
	double logarithmic time scale (right panel).}
\label{fg6}
\end{figure}

\begin{figure}
	\centering % Requires \usepackage{graphicx}
	\includegraphics[width=0.9\linewidth]{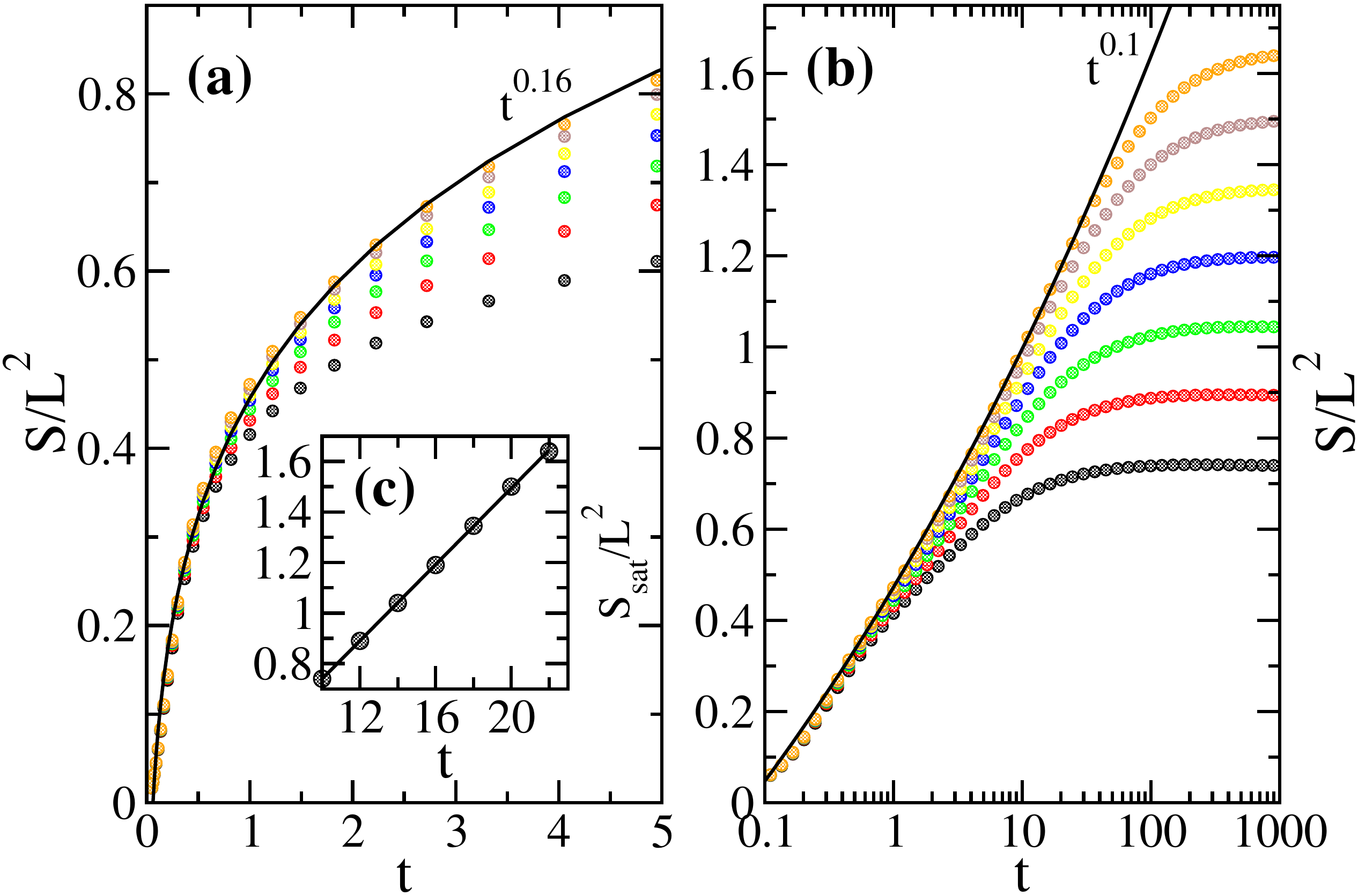} 
\caption{Entanglement entropies for $D=16$ and different system sizes $L=10,12,\cdots,22$ (from bottom to top) 
on (a) a linear, and (b) a logarithmic time scale. (c) Saturation
values. 200 to 10000 samples are used for the disorder
averages.}\label{fg7}
\end{figure}

Next, we consider the number entropy $S_N$. According to
Eq.~\eqref{bound}, we expect a scaling $S_N\sim\ln S$. This behavior is
confirmed by the numerical results, see Fig.~\ref{fg6}.
At short times, the number entropy for all disorder strengths
increases logarithmically consistent with the initial power-law
spreading of the total entanglement entropy. For disorder strengths in
the critical regime, we find $S_N\sim \ln\ln t$, see right panel of
Fig.~\ref{fg6}. We note that the observed scaling at the
metal-insulator transition in the three-dimensional Anderson model,
$S\sim \ln t$ and $S_N\sim\ln\ln t$, is exactly the same scaling which
was recently observed in the putative many-body localized (MBL) phase
of the one-dimensional Heisenberg chain
\cite{KieferUnanyan2,KieferUnanyan3}. We will discuss possible implications for MBL physics 
in interacting systems further in Sec.~\ref{Concl}.

Let us finally have a closer look at the finite-size scaling close to
the critical regime on the metallic side of the transition. In
Fig.~\ref{fg7} results for the entanglement entropy $S(t)$ for $D=16$
and different system sizes are shown.
We do see that the saturation value continues to increase with system
size. As shown in the inset of Fig.~\ref{fg7} this increase is linear
for the considered system sizes, showing that a volume law is
fulfilled and thus implying that the system is still metallic. The
time dependence is well fitted by a power law with an effective
power-law exponent which, however, is an increasing function of system
size and does depend on the fit interval. The logarithmic scaling
observed in Fig.~\ref{fg5} for $D=16$ thus holds only approximately
for $L=16$. The system is still diffusive but the convergence with
system size is slow and small systems already show approximately
critical behavior.

\begin{figure}
	\centering % Requires \usepackage{graphicx}
	\includegraphics[width=1.0\linewidth]{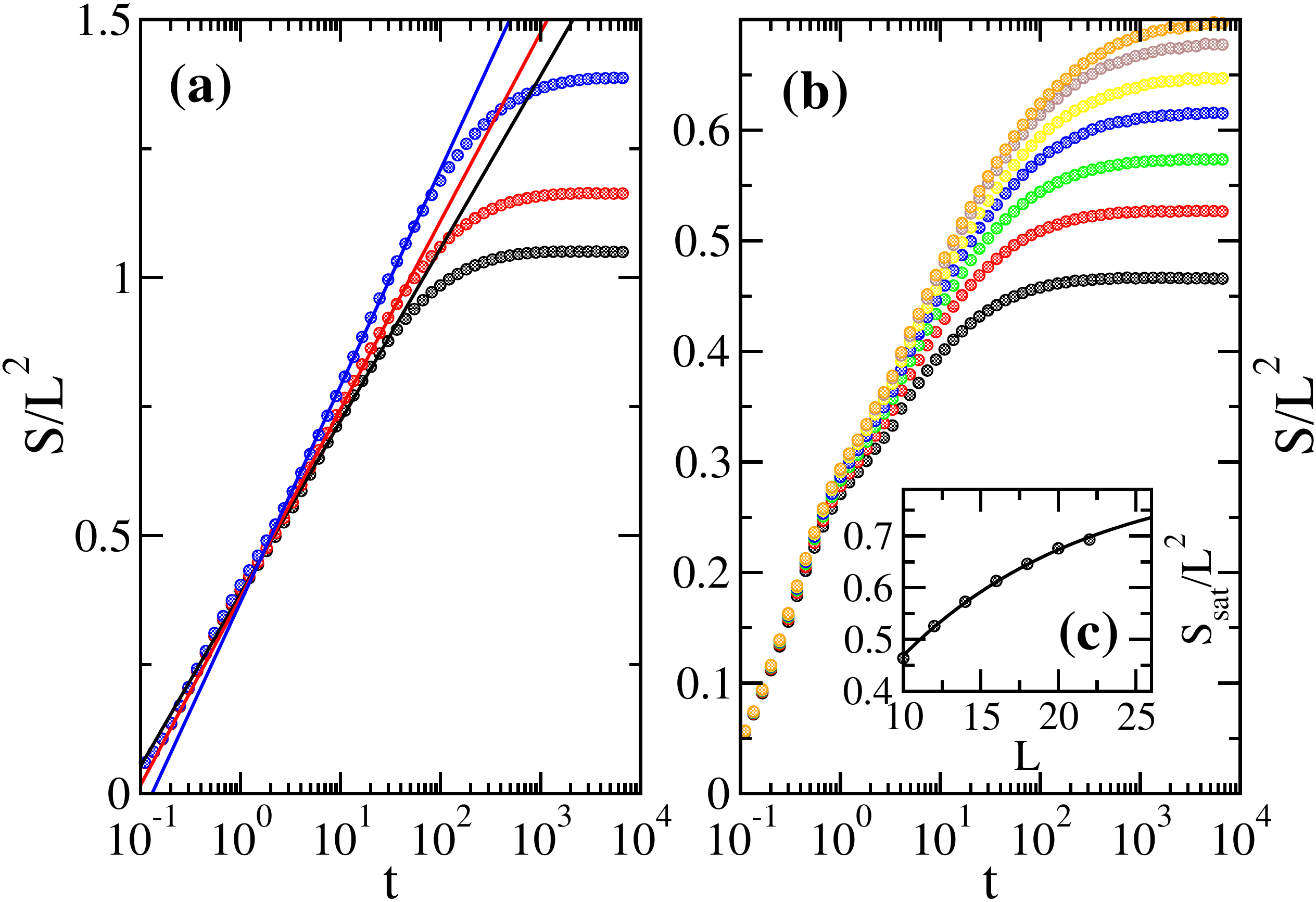} \caption{(a)
	$S(t)$ for $D=20$ and $L=16,18,22$. The lines are logarithmic
	fits. (b) $S(t)$ for $D=30$ and $L=10,12,\cdots,22$ (bottom to
	top). Averages over $200$ to $1,000$ disorder realizations
	depending on system size are shown. (c) Scaling of the
	saturation values for $D=30$ as a function of $L$ and a fit
	$S_{\textrm{sat}}/L^2=0.83(1-\exp(-0.08L))$.}\label{fg9}
\end{figure}

When we compare this with the case $D=20$ shown in Fig.~\ref{fg9}(a)
we see that the data are now consistent with $S(t)=a+\nu\ln t$ for
{\it all system sizes} shown. Only the range of time over which the
logarithmic growth is observed and the prefactor $\nu$---which
increases slowly with increasing $L$---change. Note that
for $D=20$ all eigenstates are expected to be localized, however, the
localization lengths $\xi(E)$ are much larger than the system sizes we
are able to investigate so that critical behavior is observed.

\subsection{Strong disorder}
In the regime where all eigenstates are localized, $D\gtrsim 17$, both
$S(t)$ and $S_N(t)$ will saturate in the thermodynamic limit. To
observe full localization numerically, we need $\xi\ll L$ where $L$
are the largest system sizes considered. If this is the case, then the
data will become converged in system size. As shown in
Fig.~\ref{fg9}(b), the volume law is already clearly violated for
$D=30$ although the saturation values have not quite reached the
thermodynamic limit value yet. We also note that for $D=30$ the growth of
the entanglement entropy at intermediate times is slower than
logarithmic and the scaling thus different from the critical regime.

\begin{figure}
	\centering % Requires \usepackage{graphicx}
	\includegraphics[width=1.0\linewidth]{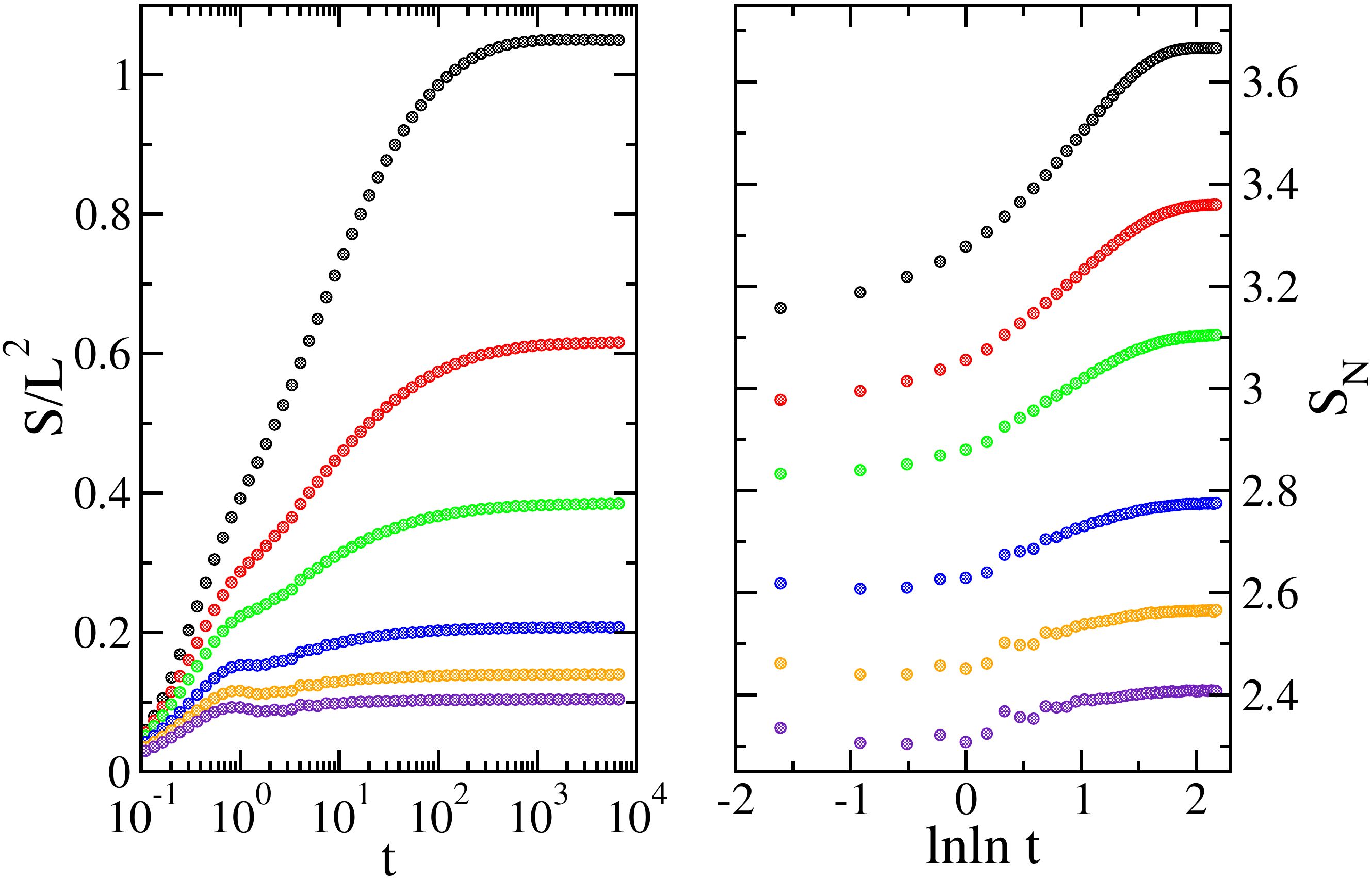} \caption{Left
	panel: $S(t)$ for $D=20,30,40,60,80,100$ (top to bottom) with
	$L=16$ fixed. Right panel: The corresponding number
	entropies.}
\label{fg8}
\end{figure}

To study the entanglement growth at intermediate times for $D>D_c$ in
more detail, we show in Fig.~\ref{fg8} data for different $D$ with the
size of the system kept fixed.
While a logarithmic scaling is visible for $D=20$ at intermediate
times, $S(t)$ for large disorder strengths starts to saturate
immediately after a fast initial increase and after going through a
local maximum. A qualitatively similar behavior is also observed for
the number entropy where an approximate $S_N\sim\ln\ln t$ scaling at
intermediate times only occurs close to the critical point while no
such scaling is observed for larger disorder. This is in contrast to
the putative ergodic-MBL transition in the disordered Heisenberg
chain. Here the entanglement entropy continues to grow as $S(t)\sim\ln
t$ for $D>D_c$ instead of saturating which is considered to be one of
the hallmarks of MBL phases. The number entropy $S_N(t)$, on the other
hand, should saturate if the system is truly localized. This has been
called into question by recent numerical results showing that
$S_N\sim\ln\ln t$ holds even for disorders much larger than the
critical value \cite{KieferUnanyan3}. One possible explanation which
has been put forward is that the observed scaling is transient and
only describes the number entropy at intermediate
times \cite{LuitzBarLev}. While the transition here is of a different
kind---with both entropies saturating for $D>D_c$---it is worth noting
that the saturation of $S_N$ is observable in exact diagonalizations
of systems with linear dimensions $L\sim 20$ in contrast to the
putative MBL transition where this is not the case.

\begin{figure}
	\centering % Requires \usepackage{graphicx}
	\includegraphics[width=1.0\linewidth]{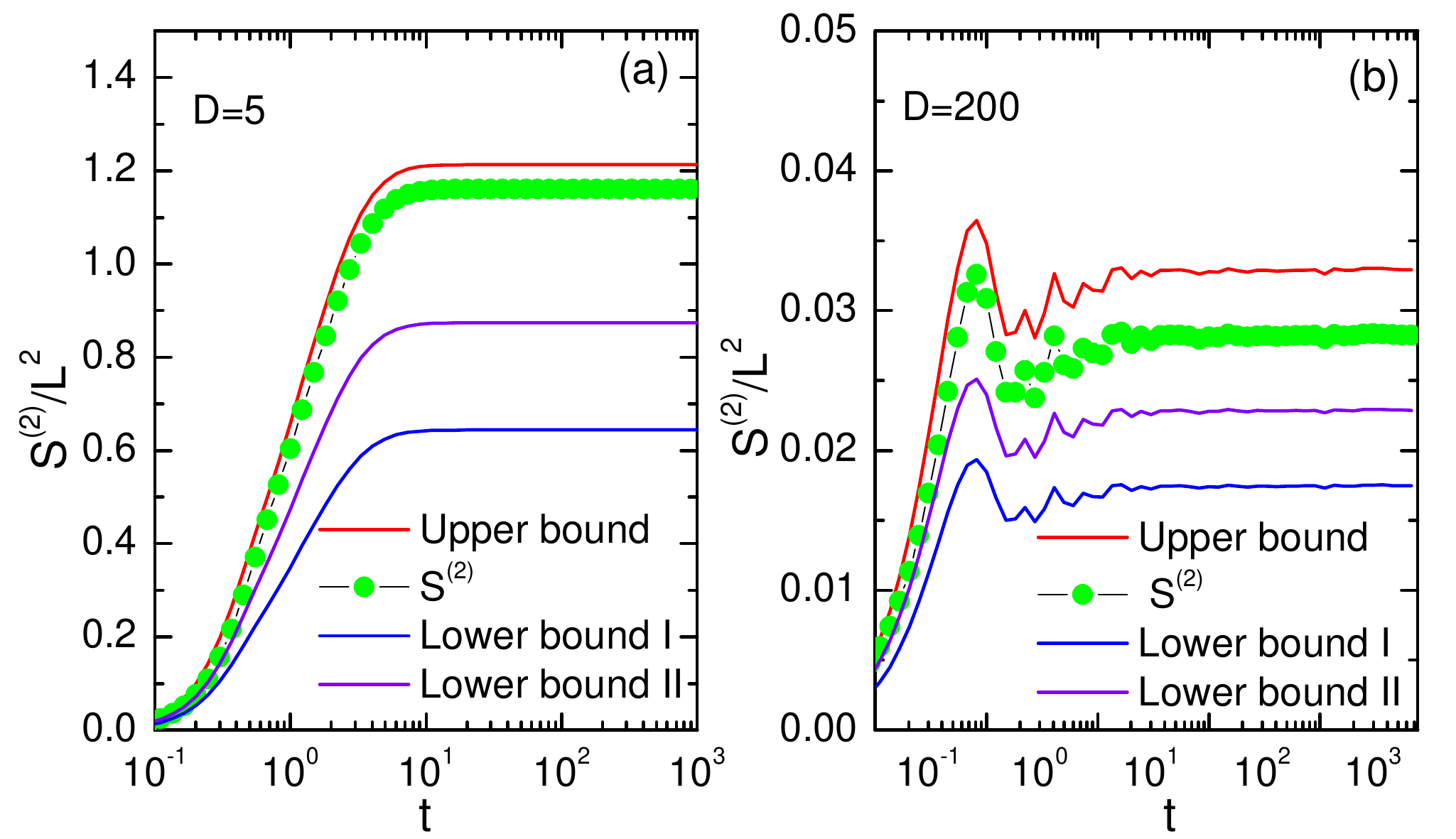} \caption{The
	second R\'enyi entropy for (a) $D=5$ and (b) $D=200$ with
	their corresponding upper and two lower
	bounds,
	$\exp(2S_N^{(2)})/(\text{e}\pi)-1/6$ (Lower bound I) and
	$\exp(2S_N)/(\text{e}\pi)-1/6$ (Lower bound II).}\label{fg10}
\end{figure}

To further stress the point that the saturation of both entanglement
and number entropies occur at the same time scale, we show in
Fig.~\ref{fg10} the bound \eqref{bound} for the second R\'enyi entropy
obtained from its corresponding number entropy.
Both in the metallic phase and in the localized phase, the bounds show
qualitatively the same time dependence as the full entropy. They do,
in particular, show saturation at the same time scale. This is exactly
the same behavior found for one-dimensional disordered Gaussian models
in Ref.~\onlinecite{KieferUnanyan1} and for the putative ergodic-MBL
transition in the disordered Heisenberg chain in
Ref.~\onlinecite{KieferUnanyan2}. 

\begin{figure}
	\centering % Requires \usepackage{graphicx}
	\includegraphics[width=1.0\linewidth]{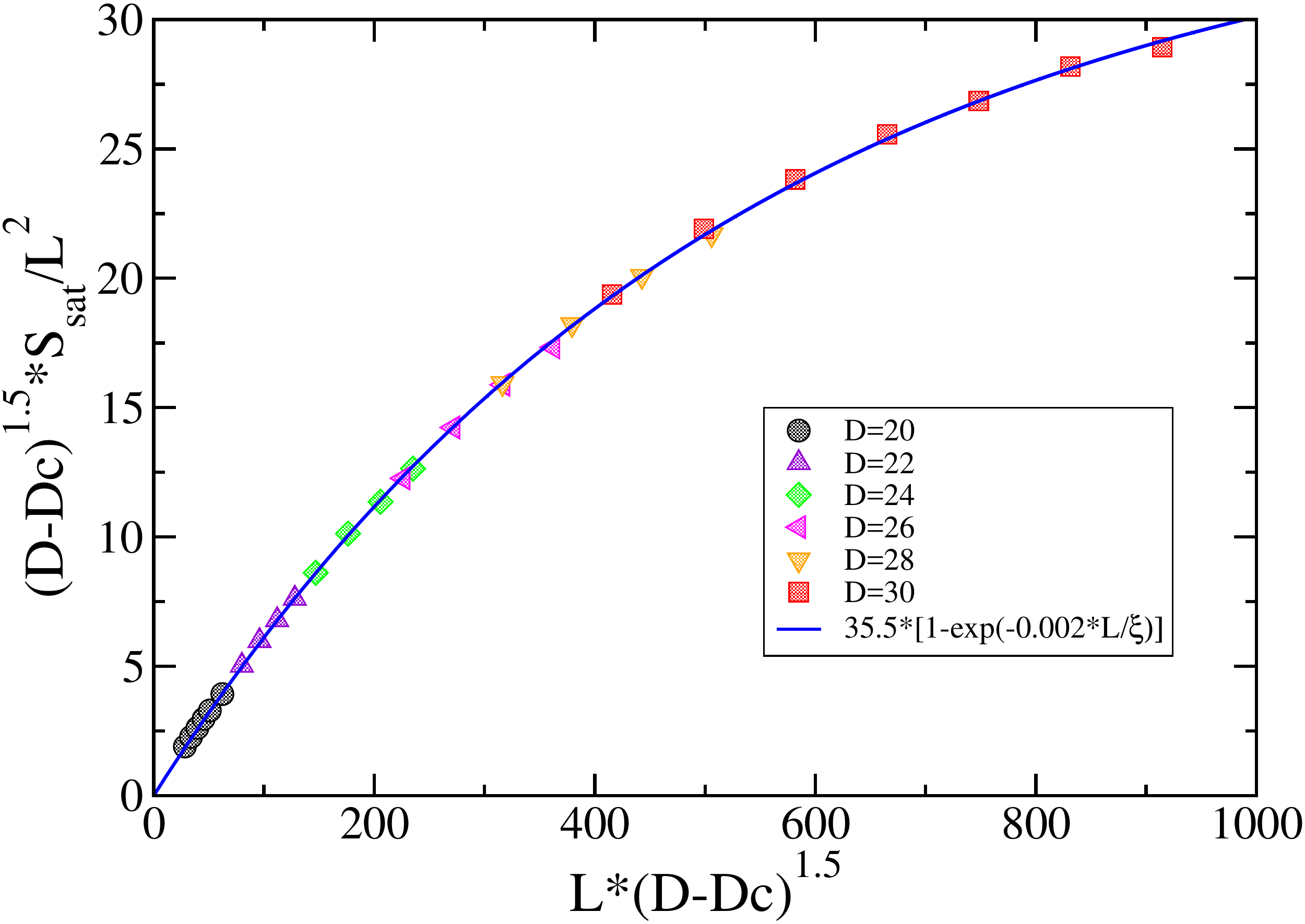}
	\caption{Scaling collapse of the saturation values
	$S_{\textrm{sat}}$ according to Eq.~\eqref{collapse} for $D_c=18$
	and $\nu=1.5$.}
\label{fgscaling}
\end{figure}

Finally, we show that our results for the entanglement entropy are
consistent with what is known about the critical properties of the
three-dimensional Anderson model at the metal-insulator transition. We
want to stress though, that studying the entanglement growth after a
quantum quench is not a suitable probe to obtain precise results for
the critical properties. In particular, this probe does not offer any
energy resolution. We therefore only try to show consistency with
previous numerical results
\cite{AndersonLocalization,BulkaKramer,SlevinMarkos,VasquezRodriguez1,VasquezRodriguez2}
using the simplest scaling form where irrelevant scaling variables are
completely ignored. In the localized phase, the saturation value in
the thermodynamic limit is given by $S_{\textrm{sat}}/L^2\sim \xi(D)$
where $\xi(D)\sim |D-D_c|^{-\nu}$ is the localization length which
diverges at the critical disorder strength $D_c$ with critical
exponent $\nu$. For a finite system, we can introduce an effective
length scale $\tilde\xi=\xi f(L/\xi)$ with $f(L/\xi)\to 1$ for
$L\to\infty$. We therefore expect the following scaling collapse in
the localized phase
\begin{equation}
\label{collapse}
\frac{S_{\textrm{sat}}}{\xi L^2} \propto f(L/\xi) \, .
\end{equation}
The two parameters controlling the scaling are: the critical disorder
strength $D_c$ and the critical exponent $\nu$. In
Fig.~\ref{fgscaling} we show that an excellent scaling collapse can be
obtained for $D_c=18$ and $\nu=1.5$.
These values are consistent with the critical disorder strength where
we expect all eigenstates to become localized and with the critical
exponent found in other studies \cite{BulkaKramer,SlevinMarkos}. We
note, however, that small variations of $D_c$ and $\nu$ are possible
without drastically affecting the scaling collapse. We have not tried
to fully optimize $D_c$ and $\nu$; the minimum appears to be quite
shallow.

\section{Conclusions}
\label{Concl}
We have studied the entanglement growth after a quantum quench
starting from an initial product state in the three-dimensional
Anderson model using exact diagonalizations. Our study has been
motivated to a large extent by the recent interest in putative
transitions from ergodic to many-body localized (MBL) phases in
interacting models such as the Heisenberg chain with random magnetic
fields. Similar to an ergodic-MBL transition, the metal-insulator
transition in the 3d Anderson model is a transition where the
entanglement properties change at some critical disorder. We find that
in the latter case the entanglement at the critical point grows as
$S(t)\sim\ln t$ while the number entropy grows as $S_N(t)\sim\ln \ln
t$. This is exactly the same behavior which has recently been found
for the entire numerically investigated part of the putative MBL phase
in the Heisenberg chain with magnetic field disorder
\cite{KieferUnanyan2,KieferUnanyan3}. This analogy supports the view 
that the MBL phase in the Heisenberg chain is not fully localized and
more akin to an extended critical phase with very slow, subdiffusive
particle transport. In contrast to the Heisenberg chain, we know for
sure that the critical behavior in the 3d Anderson model is followed
by a fully localized phase. This phase can be easily identified by
exact diagonalizations using entanglement measures as a probe for
linear system sizes which are similar to those considered in MBL
studies of interacting systems. In particular, there is no
intermediate time regime at strong disorder where $S(t)\sim \ln t$ and
$S_N(t)\sim\ln
\ln t$ still holds. Instead, for $D\gg D_c$ an initial rapid increase is
immediately followed by saturation. We have also shown that in the
localized phase of the 3d Anderson model the saturation values for
different disorder strengths and systems sizes can be collapsed onto a
single universal scaling curve leading to values for the critical
disorder strength $D_c$ and the critical exponent $\nu$ which are
consistent with previous studies. Finally, we note that the bounds for
the second R\'enyi entropy in terms of the corresponding number
entropy give a tight upper bound for small disorder while the lower
bound is approached for very strong disorder. This means that the full
entanglement can be estimated quite accurately from a measurement of
the number entropy alone. We believe this to be quite useful for
experiments on cold atomic gases where the particle number
distribution and thus the number entropy are easily accessible while a
measurement of the full entanglement entropy requires quantum
tomography which is currently only possible for very small system
sizes.

\acknowledgments
Y. Zhao and Dingyi Feng acknowledge support by Fundamental Research
Funds for the Central Universities (3102017OQD074, 310201911cx044,
31020170QD062, 3102017jghk02011) and National Natural Science
Foundation of China (NSFC) (61705185). J.~Sirker acknowledges support
by the Natural Sciences and Engineering Research Council (NSERC,
Canada) and by the Deutsche Forschungsgemeinschaft (DFG) via Research
Unit FOR 2316.

%\bibliography{/Users/js/Literatur}

\begin{thebibliography}{61}%
\makeatletter
\providecommand \@ifxundefined [1]{%
 \@ifx{#1\undefined}
}%
\providecommand \@ifnum [1]{%
 \ifnum #1\expandafter \@firstoftwo
 \else \expandafter \@secondoftwo
 \fi
}%
\providecommand \@ifx [1]{%
 \ifx #1\expandafter \@firstoftwo
 \else \expandafter \@secondoftwo
 \fi
}%
\providecommand \natexlab [1]{#1}%
\providecommand \enquote  [1]{``#1''}%
\providecommand \bibnamefont  [1]{#1}%
\providecommand \bibfnamefont [1]{#1}%
\providecommand \citenamefont [1]{#1}%
\providecommand \href@noop [0]{\@secondoftwo}%
\providecommand \href [0]{\begingroup \@sanitize@url \@href}%
\providecommand \@href[1]{\@@startlink{#1}\@@href}%
\providecommand \@@href[1]{\endgroup#1\@@endlink}%
\providecommand \@sanitize@url [0]{\catcode `\\12\catcode `\$12\catcode
  `\&12\catcode `\#12\catcode `\^12\catcode `\_12\catcode `\%12\relax}%
\providecommand \@@startlink[1]{}%
\providecommand \@@endlink[0]{}%
\providecommand \url  [0]{\begingroup\@sanitize@url \@url }%
\providecommand \@url [1]{\endgroup\@href {#1}{\urlprefix }}%
\providecommand \urlprefix  [0]{URL }%
\providecommand \Eprint [0]{\href }%
\providecommand \doibase [0]{https://doi.org/}%
\providecommand \selectlanguage [0]{\@gobble}%
\providecommand \bibinfo  [0]{\@secondoftwo}%
\providecommand \bibfield  [0]{\@secondoftwo}%
\providecommand \translation [1]{[#1]}%
\providecommand \BibitemOpen [0]{}%
\providecommand \bibitemStop [0]{}%
\providecommand \bibitemNoStop [0]{.\EOS\space}%
\providecommand \EOS [0]{\spacefactor3000\relax}%
\providecommand \BibitemShut  [1]{\csname bibitem#1\endcsname}%
\let\auto@bib@innerbib\@empty
%</preamble>
\bibitem [{\citenamefont {Calabrese}\ and\ \citenamefont
  {Cardy}(2006)}]{CalabreseCardyQuench}%
  \BibitemOpen
  \bibfield  {author} {\bibinfo {author} {\bibfnamefont {P.}~\bibnamefont
  {Calabrese}}\ and\ \bibinfo {author} {\bibfnamefont {J.}~\bibnamefont
  {Cardy}},\ }\bibfield  {title} {\bibinfo {title} {{Time dependence of
  correlation functions following a quantum quench}},\ }\href
  {https://doi.org/{10.1103/PhysRevLett.96.136801}} {\bibfield  {journal}
  {\bibinfo  {journal} {{Phys. Rev. Lett.}}\ }\textbf {\bibinfo {volume}
  {{96}}},\ \bibinfo {pages} {136801} (\bibinfo {year} {{2006}})}\BibitemShut
  {NoStop}%
\bibitem [{\citenamefont {Lieb}\ and\ \citenamefont
  {Robinson}(1972)}]{LiebRobinson}%
  \BibitemOpen
  \bibfield  {author} {\bibinfo {author} {\bibfnamefont {E.~H.}\ \bibnamefont
  {Lieb}}\ and\ \bibinfo {author} {\bibfnamefont {D.~W.}\ \bibnamefont
  {Robinson}},\ }\bibfield  {title} {\bibinfo {title} {The finite group
  velocity of quantum spin systems},\ }\href@noop {} {\bibfield  {journal}
  {\bibinfo  {journal} {Commun. Math. Phys.}\ }\textbf {\bibinfo {volume}
  {28}},\ \bibinfo {pages} {251} (\bibinfo {year} {1972})}\BibitemShut
  {NoStop}%
\bibitem [{\citenamefont {Bravyi}\ \emph {et~al.}(2006)\citenamefont {Bravyi},
  \citenamefont {Hastings},\ and\ \citenamefont {Verstraete}}]{BravyiHastings}%
  \BibitemOpen
  \bibfield  {author} {\bibinfo {author} {\bibfnamefont {S.}~\bibnamefont
  {Bravyi}}, \bibinfo {author} {\bibfnamefont {M.~B.}\ \bibnamefont
  {Hastings}},\ and\ \bibinfo {author} {\bibfnamefont {F.}~\bibnamefont
  {Verstraete}},\ }\bibfield  {title} {\bibinfo {title} {Lieb-robinson bounds
  and the generation of correlations and topological quantum order},\
  }\href@noop {} {\bibfield  {journal} {\bibinfo  {journal} {Phys. Rev. Lett.}\
  }\textbf {\bibinfo {volume} {97}},\ \bibinfo {pages} {050401} (\bibinfo
  {year} {2006})}\BibitemShut {NoStop}%
\bibitem [{\citenamefont {\ifmmode \check{Z}\else
  \v{Z}\fi{}nidari\ifmmode~\check{c}\else \v{c}\fi{}}\ \emph
  {et~al.}(2008)\citenamefont {\ifmmode \check{Z}\else
  \v{Z}\fi{}nidari\ifmmode~\check{c}\else \v{c}\fi{}}, \citenamefont {Prosen},\
  and\ \citenamefont {Prelov\ifmmode~\check{s}\else
  \v{s}\fi{}ek}}]{ZnidaricProsen}%
  \BibitemOpen
  \bibfield  {author} {\bibinfo {author} {\bibfnamefont {M.}~\bibnamefont
  {\ifmmode \check{Z}\else \v{Z}\fi{}nidari\ifmmode~\check{c}\else
  \v{c}\fi{}}}, \bibinfo {author} {\bibfnamefont {T.}~\bibnamefont {Prosen}},\
  and\ \bibinfo {author} {\bibfnamefont {P.}~\bibnamefont
  {Prelov\ifmmode~\check{s}\else \v{s}\fi{}ek}},\ }\bibfield  {title} {\bibinfo
  {title} {Many-body localization in the heisenberg $xxz$ magnet in a random
  field},\ }\href@noop {} {\bibfield  {journal} {\bibinfo  {journal} {Phys.
  Rev. B}\ }\textbf {\bibinfo {volume} {77}},\ \bibinfo {pages} {064426}
  (\bibinfo {year} {2008})}\BibitemShut {NoStop}%
\bibitem [{\citenamefont {Bardarson}\ \emph {et~al.}(2012)\citenamefont
  {Bardarson}, \citenamefont {Pollmann},\ and\ \citenamefont
  {Moore}}]{BardarsonPollmann}%
  \BibitemOpen
  \bibfield  {author} {\bibinfo {author} {\bibfnamefont {J.~H.}\ \bibnamefont
  {Bardarson}}, \bibinfo {author} {\bibfnamefont {F.}~\bibnamefont
  {Pollmann}},\ and\ \bibinfo {author} {\bibfnamefont {J.~E.}\ \bibnamefont
  {Moore}},\ }\bibfield  {title} {\bibinfo {title} {Unbounded growth of
  entanglement in models of many-body localization},\ }\href
  {https://doi.org/10.1103/PhysRevLett.109.017202} {\bibfield  {journal}
  {\bibinfo  {journal} {Phys. Rev. Lett.}\ }\textbf {\bibinfo {volume} {109}},\
  \bibinfo {pages} {017202} (\bibinfo {year} {2012})}\BibitemShut {NoStop}%
\bibitem [{\citenamefont {Prelov\ifmmode~\check{s}\else
  \v{s}\fi{}ek}(2016)}]{Prelovsek2016}%
  \BibitemOpen
  \bibfield  {author} {\bibinfo {author} {\bibfnamefont {P.}~\bibnamefont
  {Prelov\ifmmode~\check{s}\else \v{s}\fi{}ek}},\ }\bibfield  {title} {\bibinfo
  {title} {Decay of density waves in coupled one-dimensional
  many-body-localized systems},\ }\href@noop {} {\bibfield  {journal} {\bibinfo
   {journal} {Phys. Rev. B}\ }\textbf {\bibinfo {volume} {94}},\ \bibinfo
  {pages} {144204} (\bibinfo {year} {2016})}\BibitemShut {NoStop}%
\bibitem [{\citenamefont {Andraschko}\ \emph {et~al.}(2014)\citenamefont
  {Andraschko}, \citenamefont {Enss},\ and\ \citenamefont
  {Sirker}}]{AndraschkoEnssSirker}%
  \BibitemOpen
  \bibfield  {author} {\bibinfo {author} {\bibfnamefont {F.}~\bibnamefont
  {Andraschko}}, \bibinfo {author} {\bibfnamefont {T.}~\bibnamefont {Enss}},\
  and\ \bibinfo {author} {\bibfnamefont {J.}~\bibnamefont {Sirker}},\
  }\bibfield  {title} {\bibinfo {title} {Purification and many-body
  localization in cold atomic gases},\ }\href@noop {} {\bibfield  {journal}
  {\bibinfo  {journal} {Phys. Rev. Lett.}\ }\textbf {\bibinfo {volume} {113}},\
  \bibinfo {pages} {217201} (\bibinfo {year} {2014})}\BibitemShut {NoStop}%
\bibitem [{\citenamefont {Enss}\ \emph {et~al.}(2017)\citenamefont {Enss},
  \citenamefont {Andraschko},\ and\ \citenamefont
  {Sirker}}]{EnssAndraschkoSirker}%
  \BibitemOpen
  \bibfield  {author} {\bibinfo {author} {\bibfnamefont {T.}~\bibnamefont
  {Enss}}, \bibinfo {author} {\bibfnamefont {F.}~\bibnamefont {Andraschko}},\
  and\ \bibinfo {author} {\bibfnamefont {J.}~\bibnamefont {Sirker}},\
  }\bibfield  {title} {\bibinfo {title} {Many-body localization in infinite
  chains},\ }\href@noop {} {\bibfield  {journal} {\bibinfo  {journal} {Phys.
  Rev. B}\ }\textbf {\bibinfo {volume} {95}},\ \bibinfo {pages} {045121}
  (\bibinfo {year} {2017})}\BibitemShut {NoStop}%
\bibitem [{\citenamefont {Zhao}\ \emph {et~al.}(2016)\citenamefont {Zhao},
  \citenamefont {Andraschko},\ and\ \citenamefont
  {Sirker}}]{ZhaoAndraschkoSirker}%
  \BibitemOpen
  \bibfield  {author} {\bibinfo {author} {\bibfnamefont {Y.}~\bibnamefont
  {Zhao}}, \bibinfo {author} {\bibfnamefont {F.}~\bibnamefont {Andraschko}},\
  and\ \bibinfo {author} {\bibfnamefont {J.}~\bibnamefont {Sirker}},\
  }\bibfield  {title} {\bibinfo {title} {Entanglement entropy of disordered
  quantum chains following a global quench},\ }\href@noop {} {\bibfield
  {journal} {\bibinfo  {journal} {Phys. Rev. B}\ }\textbf {\bibinfo {volume}
  {93}},\ \bibinfo {pages} {205146} (\bibinfo {year} {2016})}\BibitemShut
  {NoStop}%
\bibitem [{\citenamefont {Zhao}\ \emph {et~al.}(2017)\citenamefont {Zhao},
  \citenamefont {Ahmed},\ and\ \citenamefont {Sirker}}]{ZhaoAhmedSirker}%
  \BibitemOpen
  \bibfield  {author} {\bibinfo {author} {\bibfnamefont {Y.}~\bibnamefont
  {Zhao}}, \bibinfo {author} {\bibfnamefont {S.}~\bibnamefont {Ahmed}},\ and\
  \bibinfo {author} {\bibfnamefont {J.}~\bibnamefont {Sirker}},\ }\bibfield
  {title} {\bibinfo {title} {Localization of fermions in coupled chains with
  identical disorder},\ }\href {https://doi.org/10.1103/PhysRevB.95.235152}
  {\bibfield  {journal} {\bibinfo  {journal} {Phys. Rev. B}\ }\textbf {\bibinfo
  {volume} {95}},\ \bibinfo {pages} {235152} (\bibinfo {year}
  {2017})}\BibitemShut {NoStop}%
\bibitem [{\citenamefont {Zhao}\ and\ \citenamefont
  {Sirker}(2019)}]{ZhaoSirker2019}%
  \BibitemOpen
  \bibfield  {author} {\bibinfo {author} {\bibfnamefont {Y.}~\bibnamefont
  {Zhao}}\ and\ \bibinfo {author} {\bibfnamefont {J.}~\bibnamefont {Sirker}},\
  }\bibfield  {title} {\bibinfo {title} {Logarithmic entanglement growth in
  two-dimensional disordered fermionic systems},\ }\href@noop {} {\bibfield
  {journal} {\bibinfo  {journal} {Phys. Rev. B}\ }\textbf {\bibinfo {volume}
  {100}},\ \bibinfo {pages} {014203} (\bibinfo {year} {2019})}\BibitemShut
  {NoStop}%
\bibitem [{\citenamefont {Lukin}\ \emph {et~al.}(2019)\citenamefont {Lukin},
  \citenamefont {Rispoli}, \citenamefont {Schittko}, \citenamefont {Tai},
  \citenamefont {Kaufman}, \citenamefont {Choi}, \citenamefont {Khemani},
  \citenamefont {Leonard},\ and\ \citenamefont {Greiner}}]{LukinRispoli}%
  \BibitemOpen
  \bibfield  {author} {\bibinfo {author} {\bibfnamefont {A.}~\bibnamefont
  {Lukin}}, \bibinfo {author} {\bibfnamefont {M.}~\bibnamefont {Rispoli}},
  \bibinfo {author} {\bibfnamefont {R.}~\bibnamefont {Schittko}}, \bibinfo
  {author} {\bibfnamefont {M.~E.}\ \bibnamefont {Tai}}, \bibinfo {author}
  {\bibfnamefont {A.~M.}\ \bibnamefont {Kaufman}}, \bibinfo {author}
  {\bibfnamefont {S.}~\bibnamefont {Choi}}, \bibinfo {author} {\bibfnamefont
  {V.}~\bibnamefont {Khemani}}, \bibinfo {author} {\bibfnamefont
  {J.}~\bibnamefont {Leonard}},\ and\ \bibinfo {author} {\bibfnamefont
  {M.}~\bibnamefont {Greiner}},\ }\bibfield  {title} {\bibinfo {title} {Probing
  entanglement in a many-body-localized system},\ }\href@noop {} {\bibfield
  {journal} {\bibinfo  {journal} {Science}\ }\textbf {\bibinfo {volume}
  {364}},\ \bibinfo {pages} {256} (\bibinfo {year} {2019})}\BibitemShut
  {NoStop}%
\bibitem [{\citenamefont {Anderson}(1958)}]{Anderson58}%
  \BibitemOpen
  \bibfield  {author} {\bibinfo {author} {\bibfnamefont {P.~W.}\ \bibnamefont
  {Anderson}},\ }\bibfield  {title} {\bibinfo {title} {Absence of diffusion in
  certain random lattices},\ }\href@noop {} {\bibfield  {journal} {\bibinfo
  {journal} {Phys. Rev.}\ }\textbf {\bibinfo {volume} {109}},\ \bibinfo {pages}
  {1492} (\bibinfo {year} {1958})}\BibitemShut {NoStop}%
\bibitem [{\citenamefont {Abrahams}\ \emph {et~al.}(1979)\citenamefont
  {Abrahams}, \citenamefont {Anderson}, \citenamefont {Licciardello},\ and\
  \citenamefont {Ramakrishnan}}]{AbrahamsAnderson}%
  \BibitemOpen
  \bibfield  {author} {\bibinfo {author} {\bibfnamefont {E.}~\bibnamefont
  {Abrahams}}, \bibinfo {author} {\bibfnamefont {P.~W.}\ \bibnamefont
  {Anderson}}, \bibinfo {author} {\bibfnamefont {D.~C.}\ \bibnamefont
  {Licciardello}},\ and\ \bibinfo {author} {\bibfnamefont {T.~V.}\ \bibnamefont
  {Ramakrishnan}},\ }\bibfield  {title} {\bibinfo {title} {Scaling theory of
  localization: Absence of quantum diffusion in two dimensions},\ }\href
  {https://doi.org/10.1103/PhysRevLett.42.673} {\bibfield  {journal} {\bibinfo
  {journal} {Phys. Rev. Lett.}\ }\textbf {\bibinfo {volume} {42}},\ \bibinfo
  {pages} {673} (\bibinfo {year} {1979})}\BibitemShut {NoStop}%
\bibitem [{\citenamefont {Edwards}\ and\ \citenamefont
  {Thouless}(1972)}]{EdwardsThouless}%
  \BibitemOpen
  \bibfield  {author} {\bibinfo {author} {\bibfnamefont {J.~T.}\ \bibnamefont
  {Edwards}}\ and\ \bibinfo {author} {\bibfnamefont {D.~J.}\ \bibnamefont
  {Thouless}},\ }\bibfield  {title} {\bibinfo {title} {Numerical studies of
  localization in disordered systems},\ }\href@noop {} {\bibfield  {journal}
  {\bibinfo  {journal} {J. Phys. C}\ }\textbf {\bibinfo {volume} {5}},\
  \bibinfo {pages} {807} (\bibinfo {year} {1972})}\BibitemShut {NoStop}%
\bibitem [{\citenamefont {Abrahams}(2010)}]{AndersonLocalization}%
  \BibitemOpen
  \bibinfo {editor} {\bibfnamefont {E.}~\bibnamefont {Abrahams}},\ ed.,\
  \href@noop {} {\emph {\bibinfo {title} {50 Years of Anderson Localization}}}\
  (\bibinfo  {publisher} {World Scientific},\ \bibinfo {address} {Singapore},\
  \bibinfo {year} {2010})\BibitemShut {NoStop}%
\bibitem [{\citenamefont {Igl\'oi}\ \emph {et~al.}(2012)\citenamefont
  {Igl\'oi}, \citenamefont {Szatm\'ari},\ and\ \citenamefont
  {Lin}}]{IgloiSzatmari}%
  \BibitemOpen
  \bibfield  {author} {\bibinfo {author} {\bibfnamefont {F.}~\bibnamefont
  {Igl\'oi}}, \bibinfo {author} {\bibfnamefont {Z.}~\bibnamefont
  {Szatm\'ari}},\ and\ \bibinfo {author} {\bibfnamefont {Y.-C.}\ \bibnamefont
  {Lin}},\ }\bibfield  {title} {\bibinfo {title} {Entanglement entropy dynamics
  of disordered quantum spin chains},\ }\href
  {https://doi.org/10.1103/PhysRevB.85.094417} {\bibfield  {journal} {\bibinfo
  {journal} {Phys. Rev. B}\ }\textbf {\bibinfo {volume} {85}},\ \bibinfo
  {pages} {094417} (\bibinfo {year} {2012})}\BibitemShut {NoStop}%
\bibitem [{\citenamefont {Eggarter}\ and\ \citenamefont
  {Riedinger}(1978)}]{EggarterRiedinger}%
  \BibitemOpen
  \bibfield  {author} {\bibinfo {author} {\bibfnamefont {T.~P.}\ \bibnamefont
  {Eggarter}}\ and\ \bibinfo {author} {\bibfnamefont {R.}~\bibnamefont
  {Riedinger}},\ }\bibfield  {title} {\bibinfo {title} {Singular behavior of
  tight-binding chains with off-diagonal disorder},\ }\href
  {https://doi.org/10.1103/PhysRevB.18.569} {\bibfield  {journal} {\bibinfo
  {journal} {Phys. Rev. B}\ }\textbf {\bibinfo {volume} {18}},\ \bibinfo
  {pages} {569} (\bibinfo {year} {1978})}\BibitemShut {NoStop}%
\bibitem [{\citenamefont {Balents}\ and\ \citenamefont
  {Fisher}(1997)}]{BalentsFisher97}%
  \BibitemOpen
  \bibfield  {author} {\bibinfo {author} {\bibfnamefont {L.}~\bibnamefont
  {Balents}}\ and\ \bibinfo {author} {\bibfnamefont {M.~P.~A.}\ \bibnamefont
  {Fisher}},\ }\bibfield  {title} {\bibinfo {title} {Delocalization transition
  via supersymmetry in one dimension},\ }\href
  {https://doi.org/10.1103/PhysRevB.56.12970} {\bibfield  {journal} {\bibinfo
  {journal} {Phys. Rev. B}\ }\textbf {\bibinfo {volume} {56}},\ \bibinfo
  {pages} {12970} (\bibinfo {year} {1997})}\BibitemShut {NoStop}%
\bibitem [{\citenamefont {Fisher}(1994)}]{Fisher_random}%
  \BibitemOpen
  \bibfield  {author} {\bibinfo {author} {\bibfnamefont {D.~S.}\ \bibnamefont
  {Fisher}},\ }\bibfield  {title} {\bibinfo {title} {Random antiferromagnetic
  quantum spin chains},\ }\href@noop {} {\bibfield  {journal} {\bibinfo
  {journal} {Phys. Rev. B}\ }\textbf {\bibinfo {volume} {50}},\ \bibinfo
  {pages} {3799} (\bibinfo {year} {1994})}\BibitemShut {NoStop}%
\bibitem [{\citenamefont {Fisher}(1992)}]{Fisher_random_Ising}%
  \BibitemOpen
  \bibfield  {author} {\bibinfo {author} {\bibfnamefont {D.~S.}\ \bibnamefont
  {Fisher}},\ }\bibfield  {title} {\bibinfo {title} {Random transverse field
  ising spin chains},\ }\href@noop {} {\bibfield  {journal} {\bibinfo
  {journal} {Phys. Rev. Lett.}\ }\textbf {\bibinfo {volume} {69}},\ \bibinfo
  {pages} {534} (\bibinfo {year} {1992})}\BibitemShut {NoStop}%
\bibitem [{\citenamefont {Fisher}(1995)}]{Fisher_random_Ising2}%
  \BibitemOpen
  \bibfield  {author} {\bibinfo {author} {\bibfnamefont {D.~S.}\ \bibnamefont
  {Fisher}},\ }\bibfield  {title} {\bibinfo {title} {Critical behavior of
  random transverse-field ising spin chains},\ }\href
  {https://doi.org/10.1103/PhysRevB.51.6411} {\bibfield  {journal} {\bibinfo
  {journal} {Phys. Rev. B}\ }\textbf {\bibinfo {volume} {51}},\ \bibinfo
  {pages} {6411} (\bibinfo {year} {1995})}\BibitemShut {NoStop}%
\bibitem [{\citenamefont {Klich}\ and\ \citenamefont
  {Levitov}(2008)}]{KlichLevitov}%
  \BibitemOpen
  \bibfield  {author} {\bibinfo {author} {\bibfnamefont {I.}~\bibnamefont
  {Klich}}\ and\ \bibinfo {author} {\bibfnamefont {L.~S.}\ \bibnamefont
  {Levitov}},\ }\bibfield  {title} {\bibinfo {title} {Scaling of entanglement
  entropy and superselection rules},\ }\href@noop {} {\bibfield  {journal}
  {\bibinfo  {journal} {arXiv:0812.0006}\ } (\bibinfo {year}
  {2008})}\BibitemShut {NoStop}%
\bibitem [{\citenamefont {Wiseman}\ and\ \citenamefont
  {Vaccaro}(2003)}]{WisemanVaccaro}%
  \BibitemOpen
  \bibfield  {author} {\bibinfo {author} {\bibfnamefont {H.~M.}\ \bibnamefont
  {Wiseman}}\ and\ \bibinfo {author} {\bibfnamefont {J.~A.}\ \bibnamefont
  {Vaccaro}},\ }\bibfield  {title} {\bibinfo {title} {Entanglement of
  indistinguishable particles shared between two parties},\ }\href
  {https://doi.org/10.1103/PhysRevLett.91.097902} {\bibfield  {journal}
  {\bibinfo  {journal} {Phys. Rev. Lett.}\ }\textbf {\bibinfo {volume} {91}},\
  \bibinfo {pages} {097902} (\bibinfo {year} {2003})}\BibitemShut {NoStop}%
\bibitem [{\citenamefont {Dowling}\ \emph {et~al.}(2006)\citenamefont
  {Dowling}, \citenamefont {Doherty},\ and\ \citenamefont
  {Wiseman}}]{DowlingDohertyWiseman}%
  \BibitemOpen
  \bibfield  {author} {\bibinfo {author} {\bibfnamefont {M.~R.}\ \bibnamefont
  {Dowling}}, \bibinfo {author} {\bibfnamefont {A.~C.}\ \bibnamefont
  {Doherty}},\ and\ \bibinfo {author} {\bibfnamefont {H.~M.}\ \bibnamefont
  {Wiseman}},\ }\bibfield  {title} {\bibinfo {title} {Entanglement of
  indistinguishable particles in condensed-matter physics},\ }\href
  {https://doi.org/10.1103/PhysRevA.73.052323} {\bibfield  {journal} {\bibinfo
  {journal} {Phys. Rev. A}\ }\textbf {\bibinfo {volume} {73}},\ \bibinfo
  {pages} {052323} (\bibinfo {year} {2006})}\BibitemShut {NoStop}%
\bibitem [{\citenamefont {Schuch}\ \emph
  {et~al.}(2004{\natexlab{a}})\citenamefont {Schuch}, \citenamefont
  {Verstraete},\ and\ \citenamefont {Cirac}}]{SchuchVerstraeteCirac}%
  \BibitemOpen
  \bibfield  {author} {\bibinfo {author} {\bibfnamefont {N.}~\bibnamefont
  {Schuch}}, \bibinfo {author} {\bibfnamefont {F.}~\bibnamefont {Verstraete}},\
  and\ \bibinfo {author} {\bibfnamefont {J.~I.}\ \bibnamefont {Cirac}},\
  }\bibfield  {title} {\bibinfo {title} {Nonlocal resources in the presence of
  superselection rules},\ }\href
  {https://doi.org/10.1103/PhysRevLett.92.087904} {\bibfield  {journal}
  {\bibinfo  {journal} {Phys. Rev. Lett.}\ }\textbf {\bibinfo {volume} {92}},\
  \bibinfo {pages} {087904} (\bibinfo {year} {2004}{\natexlab{a}})}\BibitemShut
  {NoStop}%
\bibitem [{\citenamefont {Schuch}\ \emph
  {et~al.}(2004{\natexlab{b}})\citenamefont {Schuch}, \citenamefont
  {Verstraete},\ and\ \citenamefont {Cirac}}]{SchuchVerstraeteCirac2}%
  \BibitemOpen
  \bibfield  {author} {\bibinfo {author} {\bibfnamefont {N.}~\bibnamefont
  {Schuch}}, \bibinfo {author} {\bibfnamefont {F.}~\bibnamefont {Verstraete}},\
  and\ \bibinfo {author} {\bibfnamefont {J.~I.}\ \bibnamefont {Cirac}},\
  }\bibfield  {title} {\bibinfo {title} {Quantum entanglement theory in the
  presence of superselection rules},\ }\href
  {https://doi.org/10.1103/PhysRevA.70.042310} {\bibfield  {journal} {\bibinfo
  {journal} {Phys. Rev. A}\ }\textbf {\bibinfo {volume} {70}},\ \bibinfo
  {pages} {042310} (\bibinfo {year} {2004}{\natexlab{b}})}\BibitemShut
  {NoStop}%
\bibitem [{\citenamefont {Song}\ \emph {et~al.}(2011)\citenamefont {Song},
  \citenamefont {Flindt}, \citenamefont {Rachel}, \citenamefont {Klich},\ and\
  \citenamefont {Le~Hur}}]{SongFlindt}%
  \BibitemOpen
  \bibfield  {author} {\bibinfo {author} {\bibfnamefont {H.~F.}\ \bibnamefont
  {Song}}, \bibinfo {author} {\bibfnamefont {C.}~\bibnamefont {Flindt}},
  \bibinfo {author} {\bibfnamefont {S.}~\bibnamefont {Rachel}}, \bibinfo
  {author} {\bibfnamefont {I.}~\bibnamefont {Klich}},\ and\ \bibinfo {author}
  {\bibfnamefont {K.}~\bibnamefont {Le~Hur}},\ }\bibfield  {title} {\bibinfo
  {title} {Entanglement entropy from charge statistics: Exact relations for
  noninteracting many-body systems},\ }\href
  {https://doi.org/10.1103/PhysRevB.83.161408} {\bibfield  {journal} {\bibinfo
  {journal} {Phys. Rev. B}\ }\textbf {\bibinfo {volume} {83}},\ \bibinfo
  {pages} {161408} (\bibinfo {year} {2011})}\BibitemShut {NoStop}%
\bibitem [{\citenamefont {Song}\ \emph {et~al.}(2012)\citenamefont {Song},
  \citenamefont {Rachel}, \citenamefont {Flindt}, \citenamefont {Klich},
  \citenamefont {Laflorencie},\ and\ \citenamefont {Le~Hur}}]{SongRachel}%
  \BibitemOpen
  \bibfield  {author} {\bibinfo {author} {\bibfnamefont {H.~F.}\ \bibnamefont
  {Song}}, \bibinfo {author} {\bibfnamefont {S.}~\bibnamefont {Rachel}},
  \bibinfo {author} {\bibfnamefont {C.}~\bibnamefont {Flindt}}, \bibinfo
  {author} {\bibfnamefont {I.}~\bibnamefont {Klich}}, \bibinfo {author}
  {\bibfnamefont {N.}~\bibnamefont {Laflorencie}},\ and\ \bibinfo {author}
  {\bibfnamefont {K.}~\bibnamefont {Le~Hur}},\ }\bibfield  {title} {\bibinfo
  {title} {Bipartite fluctuations as a probe of many-body entanglement},\
  }\href {https://doi.org/10.1103/PhysRevB.85.035409} {\bibfield  {journal}
  {\bibinfo  {journal} {Phys. Rev. B}\ }\textbf {\bibinfo {volume} {85}},\
  \bibinfo {pages} {035409} (\bibinfo {year} {2012})}\BibitemShut {NoStop}%
\bibitem [{\citenamefont {Kiefer-Emmanouilidis}\ \emph
  {et~al.}(2020{\natexlab{a}})\citenamefont {Kiefer-Emmanouilidis},
  \citenamefont {Unanyan}, \citenamefont {Sirker},\ and\ \citenamefont
  {Fleischhauer}}]{KieferUnanyan1}%
  \BibitemOpen
  \bibfield  {author} {\bibinfo {author} {\bibfnamefont {M.}~\bibnamefont
  {Kiefer-Emmanouilidis}}, \bibinfo {author} {\bibfnamefont {R.}~\bibnamefont
  {Unanyan}}, \bibinfo {author} {\bibfnamefont {J.}~\bibnamefont {Sirker}},\
  and\ \bibinfo {author} {\bibfnamefont {M.}~\bibnamefont {Fleischhauer}},\
  }\bibfield  {title} {\bibinfo {title} {Bounds on the entanglement entropy by
  the number entropy in non-interacting fermionic systems},\ }\href@noop {}
  {\bibfield  {journal} {\bibinfo  {journal} {SciPost Phys.}\ }\textbf
  {\bibinfo {volume} {8}},\ \bibinfo {pages} {083} (\bibinfo {year}
  {2020}{\natexlab{a}})}\BibitemShut {NoStop}%
\bibitem [{\citenamefont {Kiefer-Emmanouilidis}\ \emph
  {et~al.}(2020{\natexlab{b}})\citenamefont {Kiefer-Emmanouilidis},
  \citenamefont {Unanyan}, \citenamefont {Fleischhauer},\ and\ \citenamefont
  {Sirker}}]{KieferUnanyan2}%
  \BibitemOpen
  \bibfield  {author} {\bibinfo {author} {\bibfnamefont {M.}~\bibnamefont
  {Kiefer-Emmanouilidis}}, \bibinfo {author} {\bibfnamefont {R.}~\bibnamefont
  {Unanyan}}, \bibinfo {author} {\bibfnamefont {M.}~\bibnamefont
  {Fleischhauer}},\ and\ \bibinfo {author} {\bibfnamefont {J.}~\bibnamefont
  {Sirker}},\ }\bibfield  {title} {\bibinfo {title} {Evidence for unbounded
  growth of the number entropy in many-body localized phases},\ }\href@noop {}
  {\bibfield  {journal} {\bibinfo  {journal} {Phys. Rev. Lett.}\ }\textbf
  {\bibinfo {volume} {124}},\ \bibinfo {pages} {243601} (\bibinfo {year}
  {2020}{\natexlab{b}})}\BibitemShut {NoStop}%
\bibitem [{\citenamefont {Kiefer-Emmanouilidis}\ \emph
  {et~al.}(2020{\natexlab{c}})\citenamefont {Kiefer-Emmanouilidis},
  \citenamefont {Unanyan}, \citenamefont {Fleischhauer},\ and\ \citenamefont
  {Sirker}}]{KieferUnanyan3}%
  \BibitemOpen
  \bibfield  {author} {\bibinfo {author} {\bibfnamefont {M.}~\bibnamefont
  {Kiefer-Emmanouilidis}}, \bibinfo {author} {\bibfnamefont {R.}~\bibnamefont
  {Unanyan}}, \bibinfo {author} {\bibfnamefont {M.}~\bibnamefont
  {Fleischhauer}},\ and\ \bibinfo {author} {\bibfnamefont {J.}~\bibnamefont
  {Sirker}},\ }\bibfield  {title} {\bibinfo {title} {Absence of true localization in many-body localized phases},\ }\href@noop {}
  {\bibfield  {journal} {\bibinfo  {journal} {arXiv:2010.00565}\ } (\bibinfo {year}
  {2020}{\natexlab{c}})}\BibitemShut {NoStop}%
\bibitem [{\citenamefont {Bonsignori}\ \emph {et~al.}(2019)\citenamefont
  {Bonsignori}, \citenamefont {Ruggiero},\ and\ \citenamefont
  {Calabrese}}]{Bonsignori2019}%
  \BibitemOpen
  \bibfield  {author} {\bibinfo {author} {\bibfnamefont {R.}~\bibnamefont
  {Bonsignori}}, \bibinfo {author} {\bibfnamefont {P.}~\bibnamefont
  {Ruggiero}},\ and\ \bibinfo {author} {\bibfnamefont {P.}~\bibnamefont
  {Calabrese}},\ }\bibfield  {title} {\bibinfo {title} {Symmetry resolved
  entanglement in free fermionic systems},\ }\href
  {https://doi.org/10.1088/1751-8121/ab4b77} {\bibfield  {journal} {\bibinfo
  {journal} {Journal of Physics A: Mathematical and Theoretical}\ }\textbf
  {\bibinfo {volume} {52}},\ \bibinfo {pages} {475302} (\bibinfo {year}
  {2019})}\BibitemShut {NoStop}%
\bibitem [{\citenamefont {Murciano}\ \emph
  {et~al.}(2020{\natexlab{a}})\citenamefont {Murciano}, \citenamefont
  {Giulio},\ and\ \citenamefont {Calabrese}}]{MurcianodiGiulio}%
  \BibitemOpen
  \bibfield  {author} {\bibinfo {author} {\bibfnamefont {S.}~\bibnamefont
  {Murciano}}, \bibinfo {author} {\bibfnamefont {G.~D.}\ \bibnamefont
  {Giulio}},\ and\ \bibinfo {author} {\bibfnamefont {P.}~\bibnamefont
  {Calabrese}},\ }\bibfield  {title} {\bibinfo {title} {{Symmetry resolved
  entanglement in gapped integrable systems: a corner transfer matrix
  approach}},\ }\href {https://doi.org/10.21468/SciPostPhys.8.3.046} {\bibfield
   {journal} {\bibinfo  {journal} {SciPost Phys.}\ }\textbf {\bibinfo {volume}
  {8}},\ \bibinfo {pages} {46} (\bibinfo {year}
  {2020}{\natexlab{a}})}\BibitemShut {NoStop}%
\bibitem [{\citenamefont {Murciano}\ \emph
  {et~al.}(2020{\natexlab{b}})\citenamefont {Murciano}, \citenamefont
  {Giulio},\ and\ \citenamefont {Calabrese}}]{MurcianodiGiulio2}%
  \BibitemOpen
  \bibfield  {author} {\bibinfo {author} {\bibfnamefont {S.}~\bibnamefont
  {Murciano}}, \bibinfo {author} {\bibfnamefont {G.~D.}\ \bibnamefont
  {Giulio}},\ and\ \bibinfo {author} {\bibfnamefont {P.}~\bibnamefont
  {Calabrese}},\ }\bibfield  {title} {\bibinfo {title} {Entanglement and
  symmetry resolution in two dimensional free quantum field theories},\
  }\href@noop {} {\bibfield  {journal} {\bibinfo  {journal} {JHEP}\ }\textbf
  {\bibinfo {volume} {2020}},\ \bibinfo {pages} {73}}\BibitemShut {NoStop}%
\bibitem [{\citenamefont {Basko}\ \emph {et~al.}(2006)\citenamefont {Basko},
  \citenamefont {Aleiner},\ and\ \citenamefont {Altshuler}}]{BaskoAleiner}%
  \BibitemOpen
  \bibfield  {author} {\bibinfo {author} {\bibfnamefont {D.~M.}\ \bibnamefont
  {Basko}}, \bibinfo {author} {\bibfnamefont {I.~L.}\ \bibnamefont {Aleiner}},\
  and\ \bibinfo {author} {\bibfnamefont {B.~L.}\ \bibnamefont {Altshuler}},\
  }\bibfield  {title} {\bibinfo {title} {Metal-insulator transition in a weakly
  interacting many-electron system with localized single-particle states},\
  }\href@noop {} {\bibfield  {journal} {\bibinfo  {journal} {Ann. Phys.}\ }
  (\bibinfo {year} {2006})}\BibitemShut {NoStop}%
\bibitem [{\citenamefont {Oganesyan}\ and\ \citenamefont
  {Huse}(2007)}]{OganesyanHuse}%
  \BibitemOpen
  \bibfield  {author} {\bibinfo {author} {\bibfnamefont {V.}~\bibnamefont
  {Oganesyan}}\ and\ \bibinfo {author} {\bibfnamefont {D.~A.}\ \bibnamefont
  {Huse}},\ }\bibfield  {title} {\bibinfo {title} {Localization of interacting
  fermions at high temperature},\ }\href@noop {} {\bibfield  {journal}
  {\bibinfo  {journal} {Phys. Rev. B}\ }\textbf {\bibinfo {volume} {75}},\
  \bibinfo {pages} {155111} (\bibinfo {year} {2007})}\BibitemShut {NoStop}%
\bibitem [{\citenamefont {Pal}\ and\ \citenamefont {Huse}(2010)}]{PalHuse}%
  \BibitemOpen
  \bibfield  {author} {\bibinfo {author} {\bibfnamefont {A.}~\bibnamefont
  {Pal}}\ and\ \bibinfo {author} {\bibfnamefont {D.~A.}\ \bibnamefont {Huse}},\
  }\bibfield  {title} {\bibinfo {title} {Many-body localization phase
  transition},\ }\href@noop {} {\bibfield  {journal} {\bibinfo  {journal}
  {Phys. Rev. B}\ }\textbf {\bibinfo {volume} {82}},\ \bibinfo {pages} {174411}
  (\bibinfo {year} {2010})}\BibitemShut {NoStop}%
\bibitem [{\citenamefont {Abanin}\ \emph {et~al.}(2019)\citenamefont {Abanin},
  \citenamefont {Altman}, \citenamefont {Bloch},\ and\ \citenamefont
  {Serbyn}}]{AbaninRev2019}%
  \BibitemOpen
  \bibfield  {author} {\bibinfo {author} {\bibfnamefont {D.~A.}\ \bibnamefont
  {Abanin}}, \bibinfo {author} {\bibfnamefont {E.}~\bibnamefont {Altman}},
  \bibinfo {author} {\bibfnamefont {I.}~\bibnamefont {Bloch}},\ and\ \bibinfo
  {author} {\bibfnamefont {M.}~\bibnamefont {Serbyn}},\ }\bibfield  {title}
  {\bibinfo {title} {Colloquium: Many-body localization, thermalization, and
  entanglement},\ }\href {https://doi.org/10.1103/RevModPhys.91.021001}
  {\bibfield  {journal} {\bibinfo  {journal} {Rev. Mod. Phys.}\ }\textbf
  {\bibinfo {volume} {91}},\ \bibinfo {pages} {021001} (\bibinfo {year}
  {2019})}\BibitemShut {NoStop}%
\bibitem [{\citenamefont {Nandkishore}\ and\ \citenamefont
  {Huse}(2015)}]{NandkishoreHuse}%
  \BibitemOpen
  \bibfield  {author} {\bibinfo {author} {\bibfnamefont {R.}~\bibnamefont
  {Nandkishore}}\ and\ \bibinfo {author} {\bibfnamefont {D.~A.}\ \bibnamefont
  {Huse}},\ }\bibfield  {title} {\bibinfo {title} {Many-body localization and
  thermalization in quantum statistical mechanics},\ }\href@noop {} {\bibfield
  {journal} {\bibinfo  {journal} {Annual Review of Condensed Matter Physics}\
  }\textbf {\bibinfo {volume} {6}},\ \bibinfo {pages} {15} (\bibinfo {year}
  {2015})}\BibitemShut {NoStop}%
\bibitem [{\citenamefont {Altman}\ and\ \citenamefont
  {Vosk}(2015)}]{AltmanVoskReview}%
  \BibitemOpen
  \bibfield  {author} {\bibinfo {author} {\bibfnamefont {E.}~\bibnamefont
  {Altman}}\ and\ \bibinfo {author} {\bibfnamefont {R.}~\bibnamefont {Vosk}},\
  }\bibfield  {title} {\bibinfo {title} {Universal dynamics and renormalization
  in many-body-localized systems},\ }\href@noop {} {\bibfield  {journal}
  {\bibinfo  {journal} {Annual Review of Condensed Matter Physics}\ }\textbf
  {\bibinfo {volume} {6}},\ \bibinfo {pages} {383} (\bibinfo {year}
  {2015})}\BibitemShut {NoStop}%
\bibitem [{\citenamefont {Luitz}\ \emph {et~al.}(2015)\citenamefont {Luitz},
  \citenamefont {Laflorencie},\ and\ \citenamefont {Alet}}]{Luitz1}%
  \BibitemOpen
  \bibfield  {author} {\bibinfo {author} {\bibfnamefont {D.~J.}\ \bibnamefont
  {Luitz}}, \bibinfo {author} {\bibfnamefont {N.}~\bibnamefont {Laflorencie}},\
  and\ \bibinfo {author} {\bibfnamefont {F.}~\bibnamefont {Alet}},\ }\bibfield
  {title} {\bibinfo {title} {Many-body localization edge in the random-field
  heisenberg chain},\ }\href@noop {} {\bibfield  {journal} {\bibinfo  {journal}
  {Phys. Rev. B}\ }\textbf {\bibinfo {volume} {91}},\ \bibinfo {pages} {081103}
  (\bibinfo {year} {2015})}\BibitemShut {NoStop}%
\bibitem [{\citenamefont {Serbyn}\ \emph {et~al.}(2013)\citenamefont {Serbyn},
  \citenamefont {Papi\ifmmode~\acute{c}\else \'{c}\fi{}},\ and\ \citenamefont
  {Abanin}}]{SerbynPapic}%
  \BibitemOpen
  \bibfield  {author} {\bibinfo {author} {\bibfnamefont {M.}~\bibnamefont
  {Serbyn}}, \bibinfo {author} {\bibfnamefont {Z.}~\bibnamefont
  {Papi\ifmmode~\acute{c}\else \'{c}\fi{}}},\ and\ \bibinfo {author}
  {\bibfnamefont {D.~A.}\ \bibnamefont {Abanin}},\ }\bibfield  {title}
  {\bibinfo {title} {Local conservation laws and the structure of the many-body
  localized states},\ }\href@noop {} {\bibfield  {journal} {\bibinfo  {journal}
  {Phys. Rev. Lett.}\ }\textbf {\bibinfo {volume} {111}},\ \bibinfo {pages}
  {127201} (\bibinfo {year} {2013})}\BibitemShut {NoStop}%
\bibitem [{\citenamefont {Huse}\ \emph {et~al.}(2014)\citenamefont {Huse},
  \citenamefont {Nandkishore},\ and\ \citenamefont
  {Oganesyan}}]{HuseNandkishore}%
  \BibitemOpen
  \bibfield  {author} {\bibinfo {author} {\bibfnamefont {D.~A.}\ \bibnamefont
  {Huse}}, \bibinfo {author} {\bibfnamefont {R.}~\bibnamefont {Nandkishore}},\
  and\ \bibinfo {author} {\bibfnamefont {V.}~\bibnamefont {Oganesyan}},\
  }\bibfield  {title} {\bibinfo {title} {Phenomenology of fully
  many-body-localized systems},\ }\href@noop {} {\bibfield  {journal} {\bibinfo
   {journal} {Phys. Rev. B}\ }\textbf {\bibinfo {volume} {90}},\ \bibinfo
  {pages} {174202} (\bibinfo {year} {2014})}\BibitemShut {NoStop}%
\bibitem [{\citenamefont {Vosk}\ \emph {et~al.}(2015)\citenamefont {Vosk},
  \citenamefont {Huse},\ and\ \citenamefont {Altman}}]{VoskHusePRX}%
  \BibitemOpen
  \bibfield  {author} {\bibinfo {author} {\bibfnamefont {R.}~\bibnamefont
  {Vosk}}, \bibinfo {author} {\bibfnamefont {D.~A.}\ \bibnamefont {Huse}},\
  and\ \bibinfo {author} {\bibfnamefont {E.}~\bibnamefont {Altman}},\
  }\bibfield  {title} {\bibinfo {title} {Theory of the many-body localization
  transition in one-dimensional systems},\ }\href@noop {} {\bibfield  {journal}
  {\bibinfo  {journal} {Phys. Rev. X}\ }\textbf {\bibinfo {volume} {5}},\
  \bibinfo {pages} {031032} (\bibinfo {year} {2015})}\BibitemShut {NoStop}%
\bibitem [{\citenamefont {Potter}\ \emph {et~al.}(2015)\citenamefont {Potter},
  \citenamefont {Vasseur},\ and\ \citenamefont
  {Parameswaran}}]{PotterVasseurPRX}%
  \BibitemOpen
  \bibfield  {author} {\bibinfo {author} {\bibfnamefont {A.~C.}\ \bibnamefont
  {Potter}}, \bibinfo {author} {\bibfnamefont {R.}~\bibnamefont {Vasseur}},\
  and\ \bibinfo {author} {\bibfnamefont {S.~A.}\ \bibnamefont {Parameswaran}},\
  }\bibfield  {title} {\bibinfo {title} {Universal properties of many-body
  delocalization transitions},\ }\href@noop {} {\bibfield  {journal} {\bibinfo
  {journal} {Phys. Rev. X}\ }\textbf {\bibinfo {volume} {5}},\ \bibinfo {pages}
  {031033} (\bibinfo {year} {2015})}\BibitemShut {NoStop}%
\bibitem [{\citenamefont {Imbrie}(2016)}]{Imbrie2016}%
  \BibitemOpen
  \bibfield  {author} {\bibinfo {author} {\bibfnamefont {J.~Z.}\ \bibnamefont
  {Imbrie}},\ }\bibfield  {title} {\bibinfo {title} {Diagonalization and
  many-body localization for a disordered quantum spin chain},\ }\href@noop {}
  {\bibfield  {journal} {\bibinfo  {journal} {Phys. Rev. Lett.}\ }\textbf
  {\bibinfo {volume} {117}},\ \bibinfo {pages} {027201} (\bibinfo {year}
  {2016})}\BibitemShut {NoStop}%
\bibitem [{\citenamefont {Suntajs}\ \emph {et~al.}(2019)\citenamefont
  {Suntajs}, \citenamefont {Bonca}, \citenamefont {Prosen},\ and\ \citenamefont
  {Vidmar}}]{SuntajsBonca}%
  \BibitemOpen
  \bibfield  {author} {\bibinfo {author} {\bibfnamefont {J.}~\bibnamefont
  {Suntajs}}, \bibinfo {author} {\bibfnamefont {J.}~\bibnamefont {Bonca}},
  \bibinfo {author} {\bibfnamefont {T.}~\bibnamefont {Prosen}},\ and\ \bibinfo
  {author} {\bibfnamefont {L.}~\bibnamefont {Vidmar}},\ }\bibfield  {title}
  {\bibinfo {title} {Quantum chaos challenges many-body localization},\
  }\href@noop {} {\bibfield  {journal} {\bibinfo  {journal} {arXiv:
  1905.06345}\ } (\bibinfo {year} {2019})}\BibitemShut {NoStop}%
\bibitem [{\citenamefont {Suntajs}\ \emph {et~al.}(2020)\citenamefont
  {Suntajs}, \citenamefont {Bonca}, \citenamefont {Prosen},\ and\ \citenamefont
  {Vidmar}}]{SuntajsBonca2}%
  \BibitemOpen
  \bibfield  {author} {\bibinfo {author} {\bibfnamefont {J.}~\bibnamefont
  {Suntajs}}, \bibinfo {author} {\bibfnamefont {J.}~\bibnamefont {Bonca}},
  \bibinfo {author} {\bibfnamefont {T.}~\bibnamefont {Prosen}},\ and\ \bibinfo
  {author} {\bibfnamefont {L.}~\bibnamefont {Vidmar}},\ }\bibfield  {title}
  {\bibinfo {title} {Ergodicity breaking transition in finite disordered spin
  chains},\ }\href@noop {} {\bibfield  {journal} {\bibinfo  {journal} {arXiv:
  2004.01719}\ } (\bibinfo {year} {2020})}\BibitemShut {NoStop}%
\bibitem [{\citenamefont {Sels}\ and\ \citenamefont
  {Polkovnikov}(2020)}]{SelsPolkovnikov}%
  \BibitemOpen
  \bibfield  {author} {\bibinfo {author} {\bibfnamefont {D.}~\bibnamefont
  {Sels}}\ and\ \bibinfo {author} {\bibfnamefont {A.}~\bibnamefont
  {Polkovnikov}},\ }\bibfield  {title} {\bibinfo {title} {Dynamical obstruction
  to localization in a disordered spin chain},\ }\href@noop {} {\bibfield
  {journal} {\bibinfo  {journal} {arXiv: 2009.04501}\ } (\bibinfo {year}
  {2020})}\BibitemShut {NoStop}%
\bibitem [{\citenamefont {Luitz}\ and\ \citenamefont
  {Lev}(2020)}]{LuitzBarLev}%
  \BibitemOpen
  \bibfield  {author} {\bibinfo {author} {\bibfnamefont {D.~J.}\ \bibnamefont
  {Luitz}}\ and\ \bibinfo {author} {\bibfnamefont {Y.~B.}\ \bibnamefont
  {Lev}},\ }\bibfield  {title} {\bibinfo {title} {Absence of slow particle
  transport in the many-body localized phase},\ }\href
  {https://doi.org/10.1103/PhysRevB.102.100202} {\bibfield  {journal} {\bibinfo
   {journal} {Phys. Rev. B}\ }\textbf {\bibinfo {volume} {102}},\ \bibinfo
  {pages} {100202} (\bibinfo {year} {2020})}\BibitemShut {NoStop}%
\bibitem [{\citenamefont {Ospelkaus}\ \emph {et~al.}(2006)\citenamefont
  {Ospelkaus}, \citenamefont {Ospelkaus}, \citenamefont {Wille}, \citenamefont
  {Succo}, \citenamefont {Ernst}, \citenamefont {Sengstock},\ and\
  \citenamefont {Bongs}}]{OspelkausOspelkaus_Localization}%
  \BibitemOpen
  \bibfield  {author} {\bibinfo {author} {\bibfnamefont {S.}~\bibnamefont
  {Ospelkaus}}, \bibinfo {author} {\bibfnamefont {C.}~\bibnamefont
  {Ospelkaus}}, \bibinfo {author} {\bibfnamefont {O.}~\bibnamefont {Wille}},
  \bibinfo {author} {\bibfnamefont {M.}~\bibnamefont {Succo}}, \bibinfo
  {author} {\bibfnamefont {P.}~\bibnamefont {Ernst}}, \bibinfo {author}
  {\bibfnamefont {K.}~\bibnamefont {Sengstock}},\ and\ \bibinfo {author}
  {\bibfnamefont {K.}~\bibnamefont {Bongs}},\ }\bibfield  {title} {\bibinfo
  {title} {Localization of bosonic atoms by fermionic impurities in a
  three-dimensional optical lattice},\ }\href@noop {} {\bibfield  {journal}
  {\bibinfo  {journal} {Phys. Rev. Lett.}\ }\textbf {\bibinfo {volume} {96}},\
  \bibinfo {pages} {180403} (\bibinfo {year} {2006})}\BibitemShut {NoStop}%
\bibitem [{\citenamefont {Roati}\ \emph {et~al.}(2008)\citenamefont {Roati},
  \citenamefont {D'Errico}, \citenamefont {Fallani}, \citenamefont {Fattori},
  \citenamefont {Fort}, \citenamefont {Zaccanti}, \citenamefont {Modugno},
  \citenamefont {Modugno},\ and\ \citenamefont {Inguscio}}]{RoatiDerrico}%
  \BibitemOpen
  \bibfield  {author} {\bibinfo {author} {\bibfnamefont {G.}~\bibnamefont
  {Roati}}, \bibinfo {author} {\bibfnamefont {C.}~\bibnamefont {D'Errico}},
  \bibinfo {author} {\bibfnamefont {L.}~\bibnamefont {Fallani}}, \bibinfo
  {author} {\bibfnamefont {M.}~\bibnamefont {Fattori}}, \bibinfo {author}
  {\bibfnamefont {C.}~\bibnamefont {Fort}}, \bibinfo {author} {\bibfnamefont
  {M.}~\bibnamefont {Zaccanti}}, \bibinfo {author} {\bibfnamefont
  {G.}~\bibnamefont {Modugno}}, \bibinfo {author} {\bibfnamefont
  {M.}~\bibnamefont {Modugno}},\ and\ \bibinfo {author} {\bibfnamefont
  {M.}~\bibnamefont {Inguscio}},\ }\bibfield  {title} {\bibinfo {title}
  {Anderson localization of a non-interacting bose-einstein condensate},\
  }\href@noop {} {\bibfield  {journal} {\bibinfo  {journal} {Nature}\ }\textbf
  {\bibinfo {volume} {453}},\ \bibinfo {pages} {895} (\bibinfo {year}
  {2008})}\BibitemShut {NoStop}%
\bibitem [{\citenamefont {Billy}\ \emph {et~al.}(2008)\citenamefont {Billy},
  \citenamefont {Josse}, \citenamefont {Zuo}, \citenamefont {Bernard},
  \citenamefont {Hambrecht}, \citenamefont {Lugan}, \citenamefont {Cl\'ement},
  \citenamefont {Sanchez-Palencia}, \citenamefont {Bouyer},\ and\ \citenamefont
  {Aspect}}]{BillyJosse}%
  \BibitemOpen
  \bibfield  {author} {\bibinfo {author} {\bibfnamefont {J.}~\bibnamefont
  {Billy}}, \bibinfo {author} {\bibfnamefont {V.}~\bibnamefont {Josse}},
  \bibinfo {author} {\bibfnamefont {Z.}~\bibnamefont {Zuo}}, \bibinfo {author}
  {\bibfnamefont {A.}~\bibnamefont {Bernard}}, \bibinfo {author} {\bibfnamefont
  {B.}~\bibnamefont {Hambrecht}}, \bibinfo {author} {\bibfnamefont
  {P.}~\bibnamefont {Lugan}}, \bibinfo {author} {\bibfnamefont
  {D.}~\bibnamefont {Cl\'ement}}, \bibinfo {author} {\bibfnamefont
  {L.}~\bibnamefont {Sanchez-Palencia}}, \bibinfo {author} {\bibfnamefont
  {P.}~\bibnamefont {Bouyer}},\ and\ \bibinfo {author} {\bibfnamefont
  {A.}~\bibnamefont {Aspect}},\ }\bibfield  {title} {\bibinfo {title} {Direct
  observation of anderson localization of matter waves in a controlled
  disorder},\ }\href@noop {} {\bibfield  {journal} {\bibinfo  {journal}
  {Nature}\ }\textbf {\bibinfo {volume} {453}},\ \bibinfo {pages} {891}
  (\bibinfo {year} {2008})}\BibitemShut {NoStop}%
\bibitem [{\citenamefont {Bulka}\ \emph {et~al.}(1985)\citenamefont {Bulka},
  \citenamefont {Kramer},\ and\ \citenamefont {MacKinnon}}]{BulkaKramer}%
  \BibitemOpen
  \bibfield  {author} {\bibinfo {author} {\bibfnamefont {B.~R.}\ \bibnamefont
  {Bulka}}, \bibinfo {author} {\bibfnamefont {B.}~\bibnamefont {Kramer}},\ and\
  \bibinfo {author} {\bibfnamefont {A.}~\bibnamefont {MacKinnon}},\ }\bibfield
  {title} {\bibinfo {title} {Mobility edge in the three dimensional anderson
  model},\ }\href@noop {} {\bibfield  {journal} {\bibinfo  {journal} {Z. Phys.
  B}\ }\textbf {\bibinfo {volume} {60}},\ \bibinfo {pages} {13} (\bibinfo
  {year} {1985})}\BibitemShut {NoStop}%
\bibitem [{\citenamefont {Slevin}\ \emph {et~al.}(2001)\citenamefont {Slevin},
  \citenamefont {Marko\ifmmode~\check{s}\else \v{s}\fi{}},\ and\ \citenamefont
  {Ohtsuki}}]{SlevinMarkos}%
  \BibitemOpen
  \bibfield  {author} {\bibinfo {author} {\bibfnamefont {K.}~\bibnamefont
  {Slevin}}, \bibinfo {author} {\bibfnamefont {P.}~\bibnamefont
  {Marko\ifmmode~\check{s}\else \v{s}\fi{}}},\ and\ \bibinfo {author}
  {\bibfnamefont {T.}~\bibnamefont {Ohtsuki}},\ }\bibfield  {title} {\bibinfo
  {title} {Reconciling conductance fluctuations and the scaling theory of
  localization},\ }\href {https://doi.org/10.1103/PhysRevLett.86.3594}
  {\bibfield  {journal} {\bibinfo  {journal} {Phys. Rev. Lett.}\ }\textbf
  {\bibinfo {volume} {86}},\ \bibinfo {pages} {3594} (\bibinfo {year}
  {2001})}\BibitemShut {NoStop}%
\bibitem [{\citenamefont {Vasquez}\ \emph {et~al.}(2008)\citenamefont
  {Vasquez}, \citenamefont {Rodriguez},\ and\ \citenamefont
  {R\"omer}}]{VasquezRodriguez1}%
  \BibitemOpen
  \bibfield  {author} {\bibinfo {author} {\bibfnamefont {L.~J.}\ \bibnamefont
  {Vasquez}}, \bibinfo {author} {\bibfnamefont {A.}~\bibnamefont {Rodriguez}},\
  and\ \bibinfo {author} {\bibfnamefont {R.~A.}\ \bibnamefont {R\"omer}},\
  }\bibfield  {title} {\bibinfo {title} {Multifractal analysis of the
  metal-insulator transition in the three-dimensional anderson model. i.
  symmetry relation under typical averaging},\ }\href
  {https://doi.org/10.1103/PhysRevB.78.195106} {\bibfield  {journal} {\bibinfo
  {journal} {Phys. Rev. B}\ }\textbf {\bibinfo {volume} {78}},\ \bibinfo
  {pages} {195106} (\bibinfo {year} {2008})}\BibitemShut {NoStop}%
\bibitem [{\citenamefont {Rodriguez}\ \emph {et~al.}(2008)\citenamefont
  {Rodriguez}, \citenamefont {Vasquez},\ and\ \citenamefont
  {R\"omer}}]{VasquezRodriguez2}%
  \BibitemOpen
  \bibfield  {author} {\bibinfo {author} {\bibfnamefont {A.}~\bibnamefont
  {Rodriguez}}, \bibinfo {author} {\bibfnamefont {L.~J.}\ \bibnamefont
  {Vasquez}},\ and\ \bibinfo {author} {\bibfnamefont {R.~A.}\ \bibnamefont
  {R\"omer}},\ }\bibfield  {title} {\bibinfo {title} {Multifractal analysis of
  the metal-insulator transition in the three-dimensional anderson model. ii.
  symmetry relation under ensemble averaging},\ }\href
  {https://doi.org/10.1103/PhysRevB.78.195107} {\bibfield  {journal} {\bibinfo
  {journal} {Phys. Rev. B}\ }\textbf {\bibinfo {volume} {78}},\ \bibinfo
  {pages} {195107} (\bibinfo {year} {2008})}\BibitemShut {NoStop}%
\bibitem [{\citenamefont {Chung}\ and\ \citenamefont
  {Peschel}(2001)}]{ChungPeschel}%
  \BibitemOpen
  \bibfield  {author} {\bibinfo {author} {\bibfnamefont {M.-C.}\ \bibnamefont
  {Chung}}\ and\ \bibinfo {author} {\bibfnamefont {I.}~\bibnamefont
  {Peschel}},\ }\bibfield  {title} {\bibinfo {title} {Density-matrix spectra of
  solvable fermionic systems},\ }\href@noop {} {\bibfield  {journal} {\bibinfo
  {journal} {Phys. Rev. B}\ }\textbf {\bibinfo {volume} {64}},\ \bibinfo
  {pages} {064412} (\bibinfo {year} {2001})}\BibitemShut {NoStop}%
\bibitem [{\citenamefont {Peschel}(2004)}]{Peschel2004}%
  \BibitemOpen
  \bibfield  {author} {\bibinfo {author} {\bibfnamefont {I.}~\bibnamefont
  {Peschel}},\ }\bibfield  {title} {\bibinfo {title} {On the reduced density
  matrix for a chain of free electrons},\ }\href@noop {} {\bibfield  {journal}
  {\bibinfo  {journal} {J. Stat. Mech.}\ ,\ \bibinfo {pages} {P06004}}
  (\bibinfo {year} {2004})}\BibitemShut {NoStop}%
\bibitem [{\citenamefont {Peschel}\ and\ \citenamefont
  {Eisler}(2009)}]{PeschelEisler}%
  \BibitemOpen
  \bibfield  {author} {\bibinfo {author} {\bibfnamefont {I.}~\bibnamefont
  {Peschel}}\ and\ \bibinfo {author} {\bibfnamefont {V.}~\bibnamefont
  {Eisler}},\ }\bibfield  {title} {\bibinfo {title} {Reduced density matrices
  and entanglement entropy in free lattice models},\ }\href@noop {} {\bibfield
  {journal} {\bibinfo  {journal} {J. Phys. A}\ }\textbf {\bibinfo {volume}
  {42}},\ \bibinfo {pages} {504003} (\bibinfo {year} {2009})}\BibitemShut
  {NoStop}%
\bibitem [{\citenamefont {Jia}\ \emph {et~al.}(2008)\citenamefont {Jia},
  \citenamefont {Subramaniam}, \citenamefont {Gruzberg},\ and\ \citenamefont
  {Chakravarty}}]{JiaSubramaniam}%
  \BibitemOpen
  \bibfield  {author} {\bibinfo {author} {\bibfnamefont {X.}~\bibnamefont
  {Jia}}, \bibinfo {author} {\bibfnamefont {A.~R.}\ \bibnamefont
  {Subramaniam}}, \bibinfo {author} {\bibfnamefont {I.~A.}\ \bibnamefont
  {Gruzberg}},\ and\ \bibinfo {author} {\bibfnamefont {S.}~\bibnamefont
  {Chakravarty}},\ }\bibfield  {title} {\bibinfo {title} {Entanglement entropy
  and multifractality at localization transitions},\ }\href
  {https://doi.org/10.1103/PhysRevB.77.014208} {\bibfield  {journal} {\bibinfo
  {journal} {Phys. Rev. B}\ }\textbf {\bibinfo {volume} {77}},\ \bibinfo
  {pages} {014208} (\bibinfo {year} {2008})}\BibitemShut {NoStop}%
\end{thebibliography}
%

\end{document}